\documentclass[sigconf,natbib=true,anonymous=false,review=false]{acmart}
\usepackage{relsize}

\usepackage{hyperref}
\usepackage[htt]{hyphenat}

\usepackage{enumitem}

\usepackage{tabularx}
\usepackage{soul}
\usepackage{threeparttable} \usepackage{multirow}
\usepackage[usestackEOL]{stackengine} \usepackage{cellspace}
\usepackage{xcolor}
\usepackage{colortbl}

\usepackage{pifont}

\usepackage{makecell}

\usepackage{graphicx}
\graphicspath{ {./images/} }
\usepackage{balance}
\usepackage{mdframed}
\usepackage{changepage}
\usepackage[most]{tcolorbox}

\definecolor{lightgreen}{HTML}{86f986}
\definecolor{lightred}{HTML}{f98686}

\usepackage{mathtools}
\usepackage{booktabs}
\usepackage{amsmath}
\DeclareMathOperator*{\argmax}{arg\,max}
\usepackage{bbm}

\usepackage{algorithm}
\usepackage{algorithmicx}
\usepackage{algpseudocode}

\newcommand{\hi}[1]{\textcolor{black}{#1}}

\newcommand{\todo}[1]{\textcolor{red}{#1}}

\newcommand{\miniskip}{\vspace*{-.9\baselineskip}}
\newcommand{\shrink}{\vspace*{-.9\baselineskip}}

\AtBeginDocument{  \providecommand\BibTeX{{    \normalfont B\kern-0.5em{\scshape i\kern-0.25em b}\kern-0.8em\TeX}}}

\setcopyright{acmlicensed}
\copyrightyear{2026}
\acmYear{2026}
\acmDOI{XXXXXXX.XXXXXXX}

\acmConference[SIGIR '26]{The 49th International ACM SIGIR Conference on Research and Development in Information Retrieval}{July 20--24,
  2026}{Melbourne, Australia}
\acmBooktitle{Proceedings of the 49th International ACM SIGIR Conference on Research and Development in Information Retrieval (SIGIR '26), July 20--24, 2026, Melbourne, Australia}
\acmISBN{978-1-4503-XXXX-X/2026/07}
\acmPrice{}

\begin{document}

\title{FACE: A Fine-Grained Reference-Free Evaluator for Conversational Information Access}

\author{Hideaki Joko}
\affiliation{  \institution{Radboud University}
  \country{The Netherlands}
}
\email{hideaki.joko@ru.nl}

\author{Faegheh Hasibi}
\affiliation{  \institution{Radboud University}
  \country{The Netherlands}
}
\email{faegheh.hasibi@ru.nl}

\begin{abstract}
A systematic, reliable, and low-cost evaluation of \hi{Conversational Information Access (CIA)} systems remains an open challenge.
Existing \hi{reference-based} evaluation methods are proven insufficient for evaluating the dynamic nature of information access conversations, 
while existing LLM-based reference-free methods suffer from evaluation bias and limited generalizability.
This work proposes \textbf{FACE}: a \textbf{F}ine-grained, \textbf{A}spect-based \textbf{C}onversation \textbf{E}valuation method that provides evaluation scores for diverse turn and dialogue-level aspects of conversations.
FACE leverages beam search and bandit optimization to select optimized LLM instructions per evaluation aspect. It assigns scores to atomic information units (particles) using the selected instructions and then aggregates them into a single score.
\hi{We show that FACE achieves} strong correlation with human judgments, achieving system correlation of 0.9 outperforming state-of-the-art conversation evaluation methods by a large margin. We further demonstrate its optimized instructions are transferable across various LLMs and datasets.
Additionally, unlike existing LLM-based methods that provide single uninterpretable scores, FACE provides insights into the system performance and enables identifying and locating problems within conversations.
\vspace{0.5em}
\begin{center}
    \raisebox{-0.25em}{\includegraphics[width=1.25em,height=1.25em]{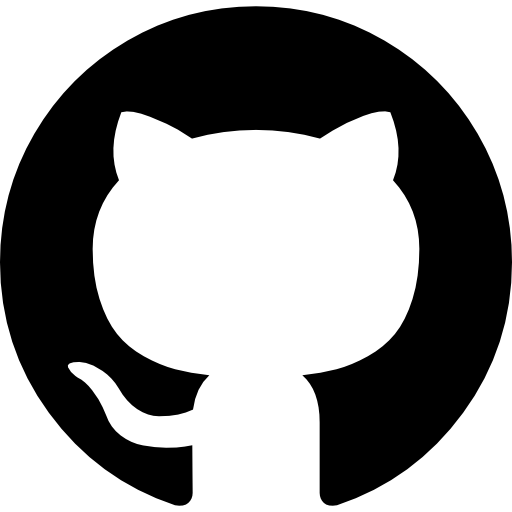}}\hspace{.3em}
    \href{https://github.com/informagi/face}{\texttt{https://github.com/informagi/face}}
\end{center}
\end{abstract}

\maketitle

\begin{figure*}[t]
\centering
\shrink
\includegraphics[width=1.0\linewidth]{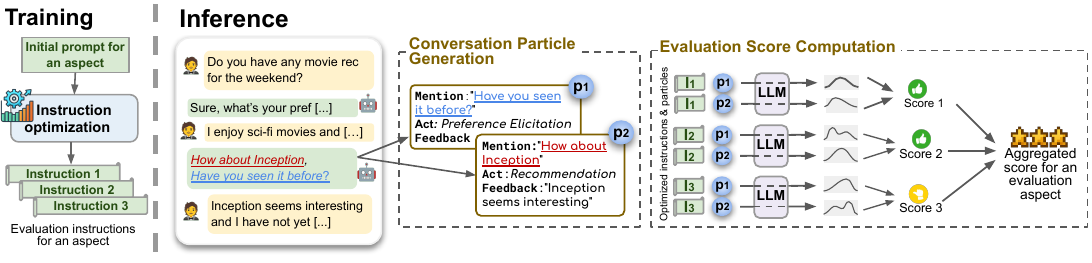}
\shrink \shrink
\caption{Illustration of FACE for a turn-level aspect. Instruction optimization generates a set of diverse evaluation instructions for the given aspect (e.g., relevance) based on an initial instruction; see example instructions in \hi{Section~\ref{sec:setup}}.
For evaluation, each conversation is decomposed into particles containing a dialogue act, an atomic statement (mention), and user feedback from the following turn. A score distribution is created for each instruction-particle pair and the weighted summation of the scores is computed. The final score is obtained by aggregating the scores across all instructions and particles.
For turn-level aspects, aggregation is performed per turn, while for dialogue-level aspects, scores of particles across the entire dialogue are aggregated.
}
\miniskip
\label{fig:method-illustration}
\end{figure*}

\section{Introduction}
\label{sec:introduction}

Recent advances in generative large language models (LLMs) have reshaped information access systems, enabling rich human-system \textit{conversational} interactions. In parallel, offline evaluation of information access systems is also evolving with the increasing adoption of LLM–based approaches~\citep{Trippas2025}. 
In industrial settings, it is common practice to design evaluation prompts for specific aspects of a system, to reduce the high cost of human evaluation, and to help early identification of system issues during the development phase~\cite{Dey:2023:DMF,Dubois:2024:LCA}.
The research community, while maintaining skepticism, has valued the potential of LLM-based evaluation, and various LLM-based evaluation methods have been proposed for information access systems~\cite{dietz:2025:llmtropes, Balog:2025:RJA, Dietz:2024:WAR, Upadhyay:2024:LSR, Farzi:2024:ELA, Thakur:2025:AST, Aliannejadi:2025:iKAT, Upadhyay:2024:LCP}.

\smallskip\noindent\textbf{Research gap.}
Most research on LLM-based evaluation for information access systems revolves around assessing the relevance of retrieved passages~\cite{Upadhyay:2024:LCP, Upadhyay:2024:LSR, Dietz:2024:WAR} or comparing generated texts against gold nuggets~\cite{Pradeep:2025:GNR, Thakur:2025:AST}. Similar efforts have been made for the offline evaluation of Conversational Information Access (CIA), where gold nuggets are generated for predefined trajectories of conversations~\cite{Abbasiantaeb:2025:CGE, Aliannejadi:2025:iKAT, Aliannejadi:2024:IKAT}. 
However, evaluation of conversational systems requires going beyond measuring factual accuracy over constrained conversational trajectories. 
\citet{Sakai:2023:SGF} identified twenty evaluation criteria (or aspects) beyond factual correctness for conversational systems (e.g., modesty and conciseness), along with a set of requirements that a conversational evaluation method should satisfy. 
To date, this broader perspective has not been fully realized. As highlighted in the latest IR strategic
meeting (SWIRL), developing ``reliable and informative methods for evaluating dynamic trajectories of conversations at scale'' is still an open question~\cite{Trippas2025}.

\smallskip\noindent\textbf{Research goal.}
In this paper, we aim to investigate ``\textit{how dynamic conversational trajectories can be evaluated beyond factual accuracy.}'' Developing such an evaluation method poses several challenges.
Reference-based evaluation methods restrict the evaluation process to predefined conversational trajectories. Even recent simulation-based approaches, such as those used in TREC iKAT 2025, remain constrained by predefined rubric questions~\cite{Aliannejadi:2025:iKAT}.
Reference-free conversational evaluation methods, on the other hand, mostly rely on LLM-based prompting and produce a single, uninterpretable score~\cite{Mehri:2020:USR,Zhong:2022:TUM,Liu:2023:GNE}, making it difficult to trace the score back to the underlying contributing factors. This violates the \textit{Specificity} requirement of conversational evaluation methods, which requires evaluation methods to ``be specific when locating problem(s) within conversations''~\citep{Sakai:2023:SGF}.

\smallskip\noindent\textbf{Method.}
We propose FACE, a \underline{F}ine-grained \underline{A}spect-based \underline{C}on\-ver\-sa\-tion \underline{E}valuation method, which is a reference-free approach for evaluating conversations across multiple evaluation aspects. 
We make the following departures from existing LLM-based conversation evaluation methods. First, we mimic the human evaluation process by modeling the diverse thought processes of multiple human assessors through a pool of optimized evaluation instructions. These optimized instructions, obtained using beam search and bandit optimization, reduce the arbitrary nature of prompt engineering and make the evaluation process robust. 
Second, we introduce the concept of ``\textit{conversation particle},'' which encompasses a \hi{composite} nugget design consisting of an atomic statement, a dialogue act, and user feedback. These conversation particles enable evaluating an atomic statement, while providing a broader view of the overall conversational state. By evaluating these conversation particles using a set of optimized instructions, our method achieves specificity. Additionally, by aggregating these scores at the turn or dialogue level, FACE supports both turn-level and dialogue-level evaluation aspects. An overall overview of FACE is illustrated in Figure~\ref{fig:method-illustration}.

\smallskip\noindent\textbf{Meta-Evaluation.}
To evaluate FACE, we collect 20,962 human annotations for 467 human-system conversations of the CRSArena-Dial dataset~\cite{Bernard:2025:CRS}. These conversations are obtained from interaction of real users with nine diverse conversational recommender systems. Human annotation of these conversations cover evaluation scores for seven evaluation aspects that are shown to be predictive of response quality and user satisfaction~\cite{Siro:2023:UPU, Siro:2022:UUS}. These include two turn-level aspects: \textit{relevance} and \textit{interestingness}, and five dialogue-level aspects: \textit{understanding}, \textit{task completion}, \textit{conversation efficiency}, \textit{interest arousal}, and \textit{overall impression}.
FACE is then applied to these evaluation aspects and the obtained evaluation scores are compared to human evaluation scores.


\smallskip\noindent\textbf{Experiments.}
Our experiments demonstrate that FACE outperforms state-of-the-art evaluation methods by a large margin, achieving system and turn/dialogue-level Spearman of 0.9 and 0.5, respectively, without observing any system-human conversations for instruction optimization.
We demonstrate that FACE provides
a reusable instruction pool that can be robustly used across different LLMs and conversation types. Remarkably, FACE outperforms
strong baselines that utilize GPT-4 as their backbone on two chit-
chat datasets, despite using an 8B-parameter Llama 3.1 as its backbone. 
Lastly, and importantly, we present a case study of two competitive conversational systems and demonstrate how FACE scores can diagnose issues and pinpoint problems within a conversation.

\smallskip\noindent\textbf{Key contributions} of this paper are summarized as follows:

\begin{itemize}[leftmargin=2em]
    \item We propose FACE, a conversation evaluation method that evaluates dynamic trajectories of conversations on diverse evaluation aspects and significantly outperforms state-of-the-art evaluation methods (Sec.~\ref{sec:results:annotation_correlation}).
    \item We introduce the notion of conversation particles and demonstrate their effectiveness in reducing bias (Sec.~\ref{sec:results:analysis}) and improving evaluation \hi{sample efficiency} (Sec.~\ref{sec:results:analysis}).
    \item We show that utilizing a pool of optimized instructions mitigates the arbitrary nature of prompt engineering for automatic evaluation and is robust across different LLMs and conversation types (Sec.~\ref{sec:results:generalizability}).
    \item We demonstrate that FACE scores are beyond a single uninterpretable score and can be used by humans to identify and locate potential system issues (\hi{Sec.~\ref{sec:results:interpretability}}).
    \item We develop the CRSArena-Eval meta-evaluation dataset, containing 20,962 human annotations over 467 conversations, for evaluating conversational evaluators. The dataset and a meta-evaluation interface for evaluating an evaluator against CRSArena-Eval are available in our repository.
    
\end{itemize}   

\section{Related Work}
\label{sec:related}

\smallskip \noindent \textbf{Automatic Conversation Evaluation.}
Although human annotations are the gold standard for evaluating CIA systems, they are expensive and time-consuming. Therefore, automatic evaluation methods have been proposed to scale up the evaluation process.
There are two main types of automatic evaluation methods: \textit{reference-based} and \textit{reference-free}~\cite{soudani2024survey}.

Reference-based methods use gold references to evaluate system responses, which include Recall@K, BLEU~\cite{Papineni:2002:BMA}, ROUGE~\cite{Lin:2004:RPA}, and BERTScore~\cite{Zhang:2020:BET}.
While these methods are effective for machine translation and question-answering, they have limitations in conversation evaluation, as they overlook diverse response possibilities and various evaluation aspects.
These limitations are supported by various studies showing a weak correlation between reference-based methods and human evaluations~\cite{Mehri:2020:USR,Bernard:2025:CRS,Liu:2016:HNE,Bernard:2025:LCE}.

Reference-free methods have been proposed to address these limitations~\cite{Mehri:2020:USR,Zhong:2022:TUM,Liu:2023:GNE}.
These methods evaluate system responses without relying on gold references, consider multiple aspects of system quality, and allow for the assessment of various response possibilities.
However, these methods primarily focus on turn-level evaluation, limiting their ability to assess the whole conversation, which are crucial for evaluating system performance in real-world scenarios~\cite{Siro:2022:UUS,Siro:2023:UPU}.
FACE addresses these limitations by evaluating the system based on the whole conversation and capturing multiple conversation trajectories from diverse user-system interactions.

\smallskip \noindent \textbf{LLMs for Evaluation.}
Recent advancements in large language models (LLMs) have led to various automatic evaluation methods, \hi{ranging from simple prompting to user simulation}, that show strong performance~\cite{Upadhyay:2024:LSR,Dubois:2024:LCA,Liu:2023:GNE,Lin:2023:UMA,Zheng:2024:JLM,Dey:2023:DMF,Farzi:2024:ELA,Aliannejadi:2025:iKAT,Thakur:2025:AST}.
However, improved performance of LLM-based methods comes with challenges.
\emph{Evaluation bias} is one of them; LLMs tend to favor longer responses (length bias)~\cite{Wang:2024:HCE,Dubois:2024:LCA} and are biased toward texts from similar models (self-bias)~\cite{Xu:2024:PPL,Liu:2023:GNE,Balog:2025:RJA}.
\emph{Generalizability} is another challenge: LLMs are sensitive to handcrafted, arbitrary prompts, which are not reusable for different models~\cite{Sclar:2024:QLM,Razavi:2025:BPS}.
FACE mitigates these biases with conversation particles and instruction optimization (see Sec.~\ref{sec:results:analysis}).

\smallskip \noindent \textbf{Instruction optimization.}
To address the arbitrary nature of handcrafted prompts, various instruction optimization (or prompt optimization) methods have been proposed~\cite{Chen:2024:IEI,Kong:2024:PPR,Pryzant:2023:APO,Ye:2024:PEP,Chen:2024:RPA,Fernando:2025:PSS,Zhou:2023:LLM,Yang:2024:LLM,Yuksekgonul:2024:TAD,Yang:2025:LLM}.
\citet{Zhou:2023:LLM} introduced APE, an automatic prompt engineering method that uses LLMs to generate prompt candidates, and perform a Monte Carlo search to find the optimal prompt.
\citet{Kong:2024:PPR} proposed PRewrite, where prompts are optimized using proximal policy optimization (PPO)~\cite{Schulman:2017:PPO}.
\citet{Pryzant:2023:APO} proposed an instruction optimization method using ``textual gradient,'' which provides natural language feedback for an LLM to optimize prompts.
These studies focus on common tasks like QA and classification, leaving conversation evaluation unexplored.
More importantly, optimized prompts are not generalizable between LLMs~\cite{Zhou:2023:LLM}, which is crucial for evaluation tasks, where reusability is essential.
In this work, we present a method for applying a textual gradient approach for conversation evaluation alongside effective strategies for transferring optimized prompts across different settings.

\smallskip \noindent \textbf{Nugget-based Evaluation.}
To assign granular evaluation scores, multiple nugget-based evaluation approaches are proposed~\cite{Voorhees:2003:OTQ,Mayfield:2024:EMR,Pradeep:2025:GNR,Yu:2025:BPS,Lin:2005:AEA,Ekstrand:2013:ESN,Takehi:2023:ODQ,Rajput:2011:NTC,Dietz:2024:WAR}.
Nugget-based evaluation was proposed~\cite{Voorhees:2003:OTQ} to assign granular scores to system responses for non-binary queries, wherein a nugget, which is an atomic piece of information, serves as the unit of evaluation, enabling a more traceable assessment.
Although the original nugget-based evaluation was intended as a manual method, many efforts have aimed to automate or semi-automate it~\cite{Mayfield:2024:EMR,Pradeep:2025:GNR,Lin:2005:AEA,Ekstrand:2013:ESN,Takehi:2023:ODQ,Rajput:2011:NTC,Dietz:2024:WAR,Abbasiantaeb:2025:CGE,Joko:2026:WIA}.
One of the earlier methods is POURPRE~\cite{Lin:2005:AEA}, an automatic nugget-based evaluation method that uses n-gram co-occurrences to assess nugget presence in system responses.
More recent research employs LLMs for nugget matching; \citet{Pradeep:2025:GNR} introduced the AutoNuggetizer framework in TREC RAG 2024, where LLMs automatically create nuggets and assign them to system responses.
However, these works are reference-based and/or focus on individual responses, and more importantly, they do not target information-seeking conversations.

\citet{Sakai:2023:SGF} proposes a conceptual framework to evaluate CIA systems by decomposing user-system interactions into units and assessing them on various aspects. While the concept is promising, its execution remains an open question, specifically how to automatically create and assess nuggets while ensuring the method's generalizability. This work addresses these challenges.

\section{Method}
\label{sec:method}

Our Fine-Grained Aspect-based Conversation Evaluation (FACE) approach handles the one-to-many nature of conversations and provides detailed scores for an evaluation aspect.
Figure~\ref{fig:method-illustration} illustrates the FACE method. 
Using beam search and bandit algorithms, FACE first optimizes a set of instructions for a given evaluation aspect.
During evaluation, a dialogue is decomposed into conversation particles, each containing a dialogue act (e.g., ``Recommendation''), mention (e.g., ``How about Inception''), and corresponding user feedback from the user response (e.g., ``Inception seems interesting''). Each particle is independently evaluated with optimized instructions via an LLM, generating score distributions and resulting in turn/dialogue-level scores for a given aspect.

We note, without detailed elaboration, that FACE is applicable to a broad range of evaluation aspects~\citep{Sakai:2023:SGF} and conversation types (e.g., recommendation, information seeking, task-oriented, and chit-chat dialogues). In this paper, we build on existing literature and available data resources~\cite{Bernard:2025:CRS} and apply FACE to conversational recommender systems for seven widely recognized turn- and dialogue-level evaluation aspects, following~\citep{Siro:2022:UUS}; see Section~\ref{sec:annotations:process} for description of these aspects.


This section describes the evaluation steps of FACE: particle generation (Sec.~\ref{sec:method:particle_generation})  and evaluation score computation (Sec.~\ref{sec:method:evaluator}), followed by the instruction optimization process (Sec.~\ref{sec:method:optimizer}).

\subsection{Conversation Particle Generation}
\label{sec:method:particle_generation}
FACE sets two goals: (1) enable reference-free evaluations to address \hi{the one-to-many nature of} natural conversation evaluation, and (2) provide fine-grained scores at both turn- and dialogue-level to locate undesired system behavior within the conversation~\cite{Sakai:2023:SGF}. To achieve these, we introduce \textit{conversation particle}, a self-contained information unit decomposed from conversations.
Each particle is composed of three parts: 
(i) Dialogue \texttt{act} is the system's action associated with the particle, such as ``recommendation'' or ``preference elicitation;'' (ii) \texttt{Mention} denotes the \hi{atomic statement} within the system's response, like ``How about the movie A?''; and (iii) \texttt{Feedback} is the user's evaluative reply, for instance, ``The movie A seems interesting.''
Following~\cite{Joko:2024:DPL}, we use 5 dialogue acts: \textit{greetings}, \textit{preference elicitation}, \textit{recommendation}, \textit{goodbye}, and \textit{others}.

We instruct an LLM to decompose system responses into particles, denoted as \textit{decomposer} $\mathcal{D}$ in the rest of the paper. 
Let $r_t$ be the target system response at turn $t$, $h$ the dialogue history preceding $r_t$, and $r_{t+1}$ the user's turn following $r_t$.
The decomposer $\mathcal{D}$ maps $(h, r_t, r_{t+1})$ to conversation particles $\mathbf{P}_r$:
\begin{equation*}
    \mathbf{P}_r = \mathcal{D}(h, r_t, r_{t+1}),
\end{equation*}
where each particle $p \in \mathbf{P}_r$ is a triplet (\texttt{act}, \texttt{mention}, \texttt{feedback}).
The full particle list for a dialogue, $\mathbf{P}_d$, is the union of particles from all responses with the dialogue, i.e., $\mathbf{P}_d = \bigcup_{r \in d} \mathbf{P}_r$.

\hi{The Prompts paragraph in Section~\ref{sec:setup}} details prompts used for particle generation. It is shown that LLMs are more effective than traditional methods like dependency parsing and information extraction for decomposing texts into atomic units~\cite{Pradeep:2025:GNR, Alaofi:2024:GIR}.

\subsection{Evaluation Score Computation}
\label{sec:method:evaluator}

FACE utilizes optimized instructions to generate scores for each conversation particle. 
Formally, given a particle $p$ and the evaluation instruction $I^a$ for the aspect $a$, an LLM generates a response $r_p^a$.
To address known issues with LLM-generated scores, such as low variance and their noise~\citep{Liu:2023:GNE}, we obtain a response distribution $\{ r_{p,i}^a \}_{i=1}^n$ and compute a weighted sum over the response set:
\begin{equation}
\label{eq:particle_score}
    \mathcal{E}_{\text{particle}}(I^a, p) = \sum_{i=1}^n r_{p,i}^a P(r_{p,i}^a|I^a, p; \theta),
\end{equation}
where $P(.)$ is a probability of $r_{p,i}^a$ from an LLM parameterized by $\theta$, and $n$ is the number of sampled evaluation responses.
These particle scores are then aggregated per turn or conversation by taking their mean:
\begin{equation*}
    \mathcal{E}(I^a, \mathbf{P}_x) = \frac{1}{|\mathbf{P}_x|} \sum_{p \in \mathbf{P}_x} \mathcal{E}_{\text{particle}}(I^a, p),
\end{equation*}
where $\mathbf{P}_x$ is the set of particles for a given turn or dialogue, depending on evaluation aspect $a$; e.g., for \textit{relevance}, aggregation is performed over particles of a turn, and for \textit{task completion} aggregation is done for all particles of the dialogue.

\textbf{A unique feature of FACE} is  utilizing diverse reasoning paths for each evaluation aspect, which is obtained by selecting optimized chain-of-thought (CoT) instructions. The intuition is that evaluation requires complex reasoning, and an optimal answer can be obtained by marginalizing various thought paths~\citep{Wang:2023:SIC}.
Here, a set of top-performing optimized instructions with various CoT instructions, $\mathbf{I}^a$, are applied to particles, and the resulting scores are aggregated to obtain the final score $s^a$:
\begin{equation*}
    s^a = \operatorname{FACE}(\mathbf{I}^a, \mathbf{P}_x) =  \frac{1}{|\mathbf{I}^a|}\,\sum_{I^a \in \mathbf{I}^a} \mathcal{E}(I^a, \mathbf{P}_x).
\end{equation*}

\subsection{Instruction Optimization}
\label{sec:method:optimizer}

Instruction optimizer generates diverse optimized CoT instructions for a given evaluation aspect.
The goal is to obtain a representative set of thought processes (via CoT) for an evaluation aspect, and leverage them to evaluate unseen human-system conversations. The optimization process is performed on annotated human-human conversations to capture human thinking process and reasoning when assessing dialogue quality. 
We note, at the outset, that all the optimization and selection algorithms are performed independently for each aspect.
For notational simplicity, we shall drop superscript $a$ from aspect-related instruction and evaluation scores in this section.

To optimize instructions, we assume access to human evaluated dialogues, $\mathbf{H} = \{(x_i, l_i)\}_{i=1}^{m}$, where $x_i$ is either a turn or an entire dialogue depending on the aspect, and $l_i$ is its label.
Similarly, we assume access to an LLM $\mathcal{L}_{1}$ that generates evaluation scores, $\mathbf{S_I}=\{(x_i,\operatorname{FACE}(\mathbf{I},\mathbf{P}_{x_i}))\}_{i=1}^{m}$, where $\mathbf{P_{x_i}}$ corresponds to particles of a specific turn or conversation and $\mathbf{P} = \bigcup_{x_i} \mathbf{P_{x_i}}$ is all particles in the dialogue collection.

The optimization objective is to identify a set of optimal instructions $\mathbf{I^*}$ that maximizes the correlation between human labels and the scores generated by the automatic evaluator:
\begin{equation*}
\label{eq:objective}
  \arg\max_{\mathbf{I^*}} \mathcal{C}(\mathbf{H}, \mathbf{S}_{\mathbf{I^*}}),
\end{equation*}
where $\mathcal{C}(.)$ represents the correlation function.

\smallskip \noindent \textbf{The optimization process} employs an LLM $\mathcal{L}_{2}$ to refine instructions based on the scores generated by the evaluator LLM $\mathcal{L}_{1}$.
FACE employs a non-parametric optimization algorithm using textual gradients~\cite{Pryzant:2023:APO}.
Here, natural language ``gradients'' (as opposed to numerical gradients) are generated to describe the shortcomings of instructions.
The gradients are used to rewrite original instructions in the opposite semantic direction.
The best instructions are iteratively selected using beam search and Upper Confidence Bound (UCB) bandits, based on correlations with human judgments. The process consists of three stages.

\smallskip \noindent \textbf{(1) Textual Gradient Generation.}
This is an iterative process, where a static prompt $\nabla$ is used for generating textual gradients (lines 6-7 of Algorithm~\ref{alg:optimization}).
At each iteration $k$, the prompt $\nabla$ takes an evaluation instruction $I_{k_j}$
from the current set of instructions $\mathbf{I}_k$, its prediction score $s_{p,k_j} = \mathcal{E}_{\text{particle}}(I_{k_j}, p)$, and the corresponding human label $l_p$ for a given particle $p$.
A set of textual gradients $\mathbf{G}_{p,k_j}$ is then generated by:
\begin{equation}
    \label{eq:textgrad}
    \mathbf{G}_{p,k_j} = \mathcal{G}_\nabla(I_{k_j}, s_{p,k_j}, l_p; \alpha), 
\end{equation}
where $\mathcal{G}_\nabla(.)$ is the gradient generation function, with parameter $\alpha$ denoting the number of gradients generated per instruction-score pair.
Since human annotations are provided at the turn- or dialogue-level, $l_p$ represents human annotation for the turn or dialogue that contains the particle $p$. 

For the prompt $\nabla$, we employ reasoning templates \cite{Ye:2024:PEP}, which provide a set of items to be considered by  $\mathcal{G}_\nabla$. 
Our items include identifying inconsistencies between the predicted and human annotations, evaluating the correctness of the current task and CoT instructions, and suggesting edits to these instructions, if necessary.

\smallskip \noindent \textbf{(2) Instruction Rewriting.}
This step updates each instruction using textual gradients (lines 8-9 of Algorithm~\ref{alg:optimization}).
For each particle $p$ at iteration $k$, we use the rewriting function $\mathcal{R}_\delta$ with the prompt $\delta$, which takes the current instruction $I_{k_j}$ and gradients $\mathbf{G}_{p,k_j}$ to obtain updated instructions $\mathbf{I}'_{p,k_j}$:
\begin{equation}
    \label{eq:rewrite}
    \mathbf{I}'_{p,k_j} = \mathcal{R}_\delta(I_{k_j}, \mathbf{G}_{p,k_j}).
\end{equation}
An LLM $\mathcal{L}_{3}$ is used for the rewriting function $\mathcal{R}_\delta$ with the prompt $\delta$ guiding it to revise the current instruction, considering the provided feedback.
We note that, while theoretically two distinct LLMs $\mathcal{L}_{2}$ and $\mathcal{L}_{3}$ are used for gradient generation and instruction rewriting, the two steps can be merged into a single LLM call by concatenating $\nabla$ and $\delta$.
This halves the number of LLM calls, resulting in a significant speed up of the optimization process.

\smallskip \noindent \textbf{(3) Instruction Selection.}
The step identifies the most promising instructions for the next iteration in two stages of selecting candidate instructions using UCB bandit, and identifying the top beam based on evaluation scores on the training data. 
This step corresponds to lines 13-16 of Algorithm~\ref{alg:optimization}.

Let 
$\mathbf{I}'_k = \bigcup_{p \in \mathbf{P}} \bigcup_j \mathbf{I}'_{p,k_j}$
be the set of all rewritten instructions at the iteration $k$. The first stage identifies $b' \geq b$ promising candidate instructions $\mathbf{I}^{\text{cand}}_k$ from the rewritten instructions $\mathbf{I}'_k$ and stores them in an \textit{instruction pool}:
\begin{align}
    \mathbf{I}^{\text{cand}}_k &= \operatorname{Select}^{\text{UCB}}_{b'}(\mathbf{I}'_k), \label{eq:ucb_select} \\
    \mathbf{I}^{\text{pool}} &\leftarrow \mathbf{I}^{\text{cand}}_k \cup \mathbf{I}^{\text{pool}}. \label{eq:pool_update}
\end{align}

The second stage then evaluates these candidate instructions by computing the correlation of the generated scores with human labels using the training set. Therefore, the instructions for the next iteration $k+1$ are generated by:
\begin{align}
    \mathbf{I}_{k+1} = \argmax_{\mathbf{I}_k \subseteq \mathbf{I}^{\text{pool}}, |\mathbf{I}_k|=b}  \mathcal{C}(\mathbf{H}, \mathbf{S}_{\mathbf{I}_k}), \label{eq:instruction_selection}
\end{align}
where $\mathbf{S}_{\mathbf{I}_k}$ denotes  all predicted scores using instructions $\mathbf{I}_k$.
This process follows a beam search approach, where at each iteration we create a pool of candidates, assess their performance, and select the top ones to form the new beam for continued exploration.
Once all iterations are complete, the final optimal instructions $\mathbf{I}^*$ are selected by their correlation scores on a validation set $\mathbf{H'}$ (line 18 of Algorithm~\ref{alg:optimization}).

\begin{algorithm}[t]
\caption{Instruction optimization of FACE}
\label{alg:optimization}
\begingroup
\fontsize{9.0}{9.2}\selectfont
\begin{algorithmic}[1]
\Require{Human evaluations $\mathbf{H}$, initial instruction $I$, iterations $K$, beam width $b$, candidate size $b'$, gradient samples $\alpha$}
\State Initialize $\mathbf{I}^{\text{pool}} \gets \emptyset$ , $\mathbf{I}_1 \gets \{I\}$
\For{$k = 1,...,K$}
    \For{each instruction $I_{k_j} \in \mathbf{I}_k$}
        \For{each particle $p \in \mathbf{P}$}
            \State $s_{p,k_j} \gets \mathcal{E}_{\text{particle}}(I_{k_j}, p)$ \Comment{Get score (Eq.~\ref{eq:particle_score})}
            \State // Textual Gradient Generation
            \State $\mathbf{G}_{p,k_j} \gets \mathcal{G}_\nabla(I_{k_j}, s_{p,k_j}, l_p; \alpha)$ \Comment{Eq.~\ref{eq:textgrad}}
            \State // Instruction Rewriting
            \State $\mathbf{I}'_{p,k_j} \gets \mathcal{R}_\delta(I_{k_j}, \mathbf{G}_{p,k_j})$ \Comment{Eq.~\ref{eq:rewrite}}
        \EndFor
    \EndFor
    \State $\mathbf{I}'_k = \bigcup_{p \in \mathbf{P}} \bigcup_j \mathbf{I}'_{p,k_j}$
    \Comment{Collect all rewrites}
    
    \State $\mathbf{I}^{\text{cand}}_k = \operatorname{Select}^{\text{UCB}}_{b'}(\mathbf{I}'_k)$ \Comment{Eq.~\ref{eq:ucb_select} and Algorithm~\ref{alg:ucb}}
    \State $\mathbf{I}^{\text{pool}} \leftarrow \mathbf{I}^{\text{cand}}_k \cup \mathbf{I}^{\text{pool}}$ \Comment{Update pool (Eq.~\ref{eq:pool_update})}
    \State // Select top-b instructions
    \State $\mathbf{I}_{k+1} \gets \argmax_{\mathbf{I}_k \subseteq \mathbf{I}^{\text{pool}}, |\mathbf{I}_k|=b} \mathcal{C}(\mathbf{H}, \mathbf{S}_{\mathbf{I}_k})$ \Comment{Eq.~\ref{eq:instruction_selection}}
\EndFor
\State $\mathbf{I}^* \gets \arg\max_{\mathbf{I}^* \subseteq \mathbf{I}^{\text{pool}}} \mathcal{C}(\mathbf{H'}, \mathbf{S}_{\mathbf{I}^*})$
\State \Return $\mathbf{I}^*$
\end{algorithmic}
\endgroup
\end{algorithm}

\smallskip \noindent \textbf{The UCB selection algorithm}, denoted as $\operatorname{Select}^{\text{UCB}}_{b'}(\cdot)$, is presented in Algorithm~\ref{alg:ucb}.
For each iteration $t$, $N_t(I)$ denotes the number of evaluations of instruction $I$ on sampled particles and $Q_t(I)$ denotes its estimated correlation.
Following~\cite{Pryzant:2023:APO}, it samples a subset of particles and their corresponding human annotations, then selects the instruction that maximizes the UCB criterion $Q_t(I) + c \sqrt{\log t / N_t(I)}$, where $c$ is an exploration constant.
The selected instruction is evaluated on the sampled particles, and its estimated effectiveness is updated based on the correlation with human annotations.
After $T$ iterations, the algorithm returns the candidate instructions $\mathbf{I}^{\text{cand}}_{k}$ containing the top $b'$ instructions according to their final estimated effectiveness $Q_T$ with the $\operatorname{SelectTopInstructions}_{b'}(Q_T)$ function.
This forms a set of promising candidates for the next stage of the selection process.
We note that, for efficient execution of the UCB process, we approximate it by dividing $T$ into multiple small batches and processing each batch in parallel.

\begin{algorithm}[t]
\caption{$\operatorname{Select}^{\text{UCB}}_{b'}(\cdot)$ - Candidate Selection with UCB Bandits}
\label{alg:ucb}
\begingroup
\fontsize{9.0}{12.5}\selectfont
\begin{algorithmic}[1]
\Require{All rewritten instructions $\mathbf{I}'_k$, particles $\mathbf{P}$, human annotations $\mathbf{H}$, number of UCB iterations $T$, and the number of selected instructions $b'$.}
\State Initialize $N_t(I) \gets 0, Q_t(I) \gets 0, \forall I \in \mathbf{I}'_k$
\For{$t = 1,...,T$}
    \State // Sample particles and corresponding labels uniformly
    \State $\mathbf{P}_{\text{smp}} \subset \mathbf{P}$, $\mathbf{H}_{\text{smp}} \subset \mathbf{H}$
    \State // Select the instruction with the highest UCB criterion
    \State $I \gets \arg\max_{I \in \mathbf{I}'_k} \{Q_t(I) + c\sqrt{\frac{\log t}{N_t(I)}}\}$
    \State // Evaluate the instruction $I$ on the sampled particles
    \State $\mathbf{S}_{\text{smp}} \gets \{\mathcal{E}_{\text{particle}}(I, p)\}_{p \in \mathbf{P}_{\text{smp}}}$
    \State // Compute correlation and update UCB criterion
   \State Observe reward $r \gets \mathcal{C}(\mathbf{S}_{\text{smp}}, \mathbf{H}_{\text{smp}})$
    \State $N_t(I) \gets N_t(I) + |\mathbf{P}_{\text{smp}}|$
    \State $Q_t(I) \gets Q_t(I) + \frac{r - Q_t(I)}{N_t(I)}$
\EndFor
\State \Return $\mathbf{I}^{\text{cand}}_k \gets \operatorname{SelectTopInstructions}_{b'}(Q_T)$
\end{algorithmic}
\endgroup
\end{algorithm}

\section{Human Annotation Collection}
\label{sec:annotations}
\suppressfloats[t]

To assess the correlation of automatic conversation evaluation methods with human judgments, we develop a dataset and crowdsource human annotations on a set of human-system conversations.
\hi{While FACE method can be applicable to any type of conversation (cf. Section~\ref{sec:method}), we build on an existing dataset of diverse human-system conversations and focus on recommendation conversations as a case study.}

\subsection{Dialogue Annotation}
\label{sec:annotations:process}

\medskip \noindent \textbf{Human-System Dialogues.}
To evaluate automatic evaluation methods, we need to compute the correlation of their scores with human quality annotations of user-system conversations.
This involves collecting user interactions with a variety of systems and annotating these conversations across multiple evaluation aspects.
Similarly to~\citet{Mehri:2020:USR}, our goal in creating such a dataset is not to train and benchmark the best-performing systems, but rather to collect responses with varying quality of systems and obtain reliable human judgments of these responses.
We select the CRSArena-Dial dataset~\cite{Bernard:2025:CRS} for this purpose, which contains, to the best of our knowledge, the most recent \hi{multi-turn dialogue dataset between human users and wide range of conversational recommender systems (CRSs), including LLM-based ones.}

The CRSArena-Dial dataset consists of human conversations with nine state-of-the-art CRSs, including KBRD~\citep{Chen:2019:KBR}, BARCOR~\citep{Wang:2022:BTU}, UniCRS~\citep{Wang:2022:UCR}, ChatGPT-based CRS~\citep{Wang:2023:REC}, and CRB-CRS~\citep{Manzoor:2022:RBC}, each developed based on OpenDialKG~\citep{Moon:2019:OEC} and ReDial~\citep{Li:2018:TDC} datasets, except CRB-CRS, which is solely on the ReDial dataset.
To ensure the quality of dialogues, we excluded seven dialogues that were unsuitable for our annotation, such as those with only a single user utterance, resulting in 467 dialogues with a total of 2,235 system responses for our annotations.

\medskip \noindent \textbf{Evaluation Aspects.}
Over 50 aspects have been proposed for evaluating conversational systems in the literature.\footnote{The full list of these aspects, compiled from 21 studies, is available in the GitHub repository of the paper.}
We follow~\citet{Siro:2022:UUS} and collect annotations for seven evaluation aspects, whose effectiveness for CRS evaluation has been rigorously validated~\cite{Siro:2022:UUS,Siro:2023:UPU}.
These aspects cover both system and user-centric features of a conversation.
The aspects and their descriptions (used as instructions to annotators) are as follows:

\medskip

\smallskip

\hangindent=1em \textbf{Turn-level Aspects:}

\begin{table}[t]
\caption{Inter-annotator agreement of CRSArena-Eval and AB-ReDial~\cite{Siro:2023:UPU} based on Pearson's $r$, Spearman's $\rho$, and Krippendorff's $\alpha$ correlations.}
\shrink
\label{tab:agreement}
\setlength{\tabcolsep}{3.5pt}
\begin{tabular}{l || c c c | c c c}
\hline
\multirow{2}{*}{\textbf{Aspect}} & \multicolumn{3}{c|}{\textbf{CRSArena-Eval}} & \multicolumn{3}{c}{\textbf{AB-ReDial}} \\
 & $r$ & $\rho$ & $\alpha$ & $r$ & $\rho$ & $\alpha$ \\
\Xhline{2pt}
\multicolumn{7}{l}{\textbf{Turn-level}} \\
\hline
Relevance         & \textbf{0.613} & \textbf{0.611} & \textbf{0.612} & 0.527 & 0.502 & 0.526 \\
Interestingness   & \textbf{0.386} & \textbf{0.386} & \textbf{0.375} & 0.209 & 0.217 & 0.209 \\
\hline
\multicolumn{7}{l}{\textbf{Dialogue-level}} \\
\hline
Understanding     & \textbf{0.505} & \textbf{0.477} & \textbf{0.496} & 0.321 & 0.313 & 0.318 \\
Task Completion   & \textbf{0.481} & \textbf{0.440} & \textbf{0.482} & 0.345 & 0.321 & 0.346 \\
Efficiency        & \textbf{0.297} & \textbf{0.297} & \textbf{0.289} & 0.225 & 0.225 & 0.226 \\
Interest Arousal  & 0.242 & 0.241 & 0.226 & \textbf{0.254} & \textbf{0.291} & \textbf{0.247} \\
Overall Impression & \textbf{0.573} & \textbf{0.526} & \textbf{0.572} & 0.321 & 0.300 & 0.321 \\
\hline
\end{tabular}
\end{table}

\hangindent=2em \textit{Relevance (0--3):} Does the assistant's response make sense and meet the user's interests?

\hangindent=2em \textit{Interestingness (0--2):} Does the response make the user want to continue the conversation?

\textbf{Dialogue-level Aspects:}

\hangindent=2em \textit{Understanding (0--2):} Does the assistant understand the user's request and try to fulfill it?

\hangindent=2em \textit{Task Completion (0--2):} Does the assistant make suggestions that the user finally accepts?

\hangindent=2em \textit{Efficiency (0--1):} Does the assistant suggest items matching the user's interests within the first three interactions?

\hangindent=2em \textit{Interest Arousal (0--2):} Does the assistant try to spark the user's interest in something new?

\hangindent=2em \textit{Overall Impression (0--4):} What is the overall impression of the assistant's performance?

\smallskip
\noindent
These instructions and scales are based on~\cite{Siro:2022:UUS}, with minor adjustments from~\citet{Sakai:2023:SGF} for clarity. Turn-level aspects are evaluated for each system turn, while dialogue-level aspects are evaluated for the entire dialogue.

Although we select seven thoroughly assessed aspects supported by the existing literature~\cite{Siro:2023:UPU}, 
we note that FACE can be applied to a broad range of aspects and the choice of aspects is often context-dependent.
Therefore, FACE algorithm allows researchers and practitioners to explore diverse aspects relevant to their own systems.

\begin{table*}[t]
\centering
\caption{Annotation correlations  for reference-free evaluation methods. 
All columns show the correlations averaged over all aspects.
Rel., Int., Und., Task, Eff., Int, and Overall denote relevance, interestingness, understanding, task completion, efficiency, interest arousal, and overall impression, respectively.
All FACE correlations are statistically significant with $p < 0.01$.}
\shrink
\label{tab:correlation_main}
\setlength{\tabcolsep}{3.5pt}
\begin{tabular}{l||cccc|cccccccccc|cc}
\hline
\multirow{3}{*}{\textbf{Methods}} & 
\multicolumn{4}{c|}{\textbf{Turn-level}} & 
\multicolumn{10}{c|}{\textbf{Dialogue-level}} & \multicolumn{2}{c}{\multirow{2}{*}{\textbf{All}}}\\
& \multicolumn{2}{c}{\textbf{Rel.}} & \multicolumn{2}{c|}{\textbf{Int.}} & 
\multicolumn{2}{c}{\textbf{Und.}} & \multicolumn{2}{c}{\textbf{Task}} & 
\multicolumn{2}{c}{\textbf{Eff.}} & \multicolumn{2}{c}{\textbf{Int.}} & 
\multicolumn{2}{c|}{\textbf{Overall}} & & \\
 & $r$ & $\rho$ & $r$ & $\rho$ & $r$ & $\rho$ & $r$ & $\rho$ & $r$ & $\rho$ & $r$ & $\rho$ & $r$ & $\rho$ & $r$ & $\rho$\\
\Xhline{2pt}
\multicolumn{17}{c}{\textbf{CRSArena-Eval (RD)}} \\
\hline
LLM\textsuperscript{Direct} & 0.464 & 0.455 & 0.248 & 0.260 & 0.522 & 0.482 & 0.405 & 0.363 & 0.101 & 0.101 & 0.217 & 0.203 & 0.564 & 0.522 & 0.360 & 0.341\\
LLM\textsuperscript{CoT+ICL} & 0.453 & 0.446 & 0.175 & 0.177 & 0.481 & 0.457 & 0.425 & 0.400 & 0.174 & 0.174 & 0.188 & 0.174 & 0.498 & 0.472 & 0.342 & 0.329\\
UniEval & 0.311 & 0.288 & 0.182 & 0.242 & 0.246 & 0.225 & -- & -- & -- & -- & -- & -- & 0.395 & 0.387 & -- & --\\
G-Eval & 0.490 & 0.471 & 0.302 & 0.289 & 0.490 & 0.444 & 0.351 & 0.364 & \textbf{0.488} & 0.482 & 0.332 & 0.325 & 0.577 & 0.577 & 0.433 & 0.422 \\
\hline
FACE w/o train & 0.468 & 0.462 & 0.279 & 0.290 & 0.605 & 0.574 & 0.482 & 0.392 & 0.339 & 0.423 & 0.235 & 0.255 & 0.617 & 0.555 & 0.432 & 0.422 \\
FACE & \textbf{0.549} & \textbf{0.550} & \textbf{0.443} & \textbf{0.437} & \textbf{0.650} & \textbf{0.635} & \textbf{0.570} & \textbf{0.453} & 0.484 & \textbf{0.534} & \textbf{0.447} & \textbf{0.430} & \textbf{0.712} & \textbf{0.668} & \textbf{0.551} & \textbf{0.530} \\
\hline
\multicolumn{17}{c}{\textbf{CRSArena-Eval (KG)}} \\
\hline
LLM\textsuperscript{Direct} & 0.452 & 0.452 & 0.238 & 0.231 & 0.599 & 0.546 & 0.538 & 0.481 & 0.137 & 0.137 & 0.425 & 0.378 & 0.655 & 0.557 & 0.435 & 0.397 \\
LLM\textsuperscript{CoT+ICL} & 0.419 & 0.408 & 0.203 & 0.190 & 0.562 & 0.520 & 0.475 & 0.434 & 0.114 & 0.114 & 0.309 & 0.279 & 0.599 & 0.521 & 0.383 & 0.352 \\
UniEval & 0.416 & 0.428 & 0.262 & 0.401 & 0.563 & 0.541 & -- & -- & -- & -- & -- & -- & 0.618 & 0.659 & -- & -- \\
G-Eval & 0.533 & 0.505 & 0.334 & 0.316 & 0.535 & 0.475 & 0.430 & 0.422 & 0.485 & 0.463 & 0.424 & 0.403 & 0.656 & 0.642 & 0.485 & 0.461 \\
\hline
FACE w/o train & 0.492 & 0.486 & 0.308 & 0.324 & 0.664 & 0.611 & 0.426 & 0.411 & 0.240 & 0.419 & 0.297 & 0.322 & 0.672 & 0.557 & 0.443 & 0.447 \\
FACE & \textbf{0.543} & \textbf{0.527} & \textbf{0.471} & \textbf{0.453} & \textbf{0.719} & \textbf{0.677} & \textbf{0.593} & \textbf{0.484} & \textbf{0.518} & \textbf{0.543} & \textbf{0.449} & \textbf{0.404} & \textbf{0.766} & \textbf{0.679} & \textbf{0.580} & \textbf{0.538} \\
\hline
\end{tabular}
\end{table*}

\smallskip \noindent \textbf{Annotation Interface.}
To crowdsource high-quality annotations, we built an interface, through multiple pilot experiments, to overcome the widely reported challenges in the literature~\cite{Joko:2024:DPL,Bernard:2023:MGS,Joko:2021:CEL,Eickhoff:2011:HCY}. This includes workers
(1) skipping reading context, (2) misunderstanding aspect definitions, (3) getting distracted by overwhelming information on the annotation page, and (4) annotating randomly without focus.
Our interface requires users to pass a quiz on aspect definitions and enforces the annotation of each turn before evaluating the entire dialogue. It further includes hidden tests with expert-verified answers, and workers who fail to meet the required agreement are dismissed.

\smallskip \noindent \textbf{Participants and Quality Control.}
We recruited Prolific\footnote{\url{https://www.prolific.co/}} workers from English-speaking countries with a $ 100\% $ approval rate and $ \geq 1000 $ previous submissions.
Considering some workers exhibit behavior aimed at just maximizing financial gain~\cite{Eickhoff:2011:HCY}, we filtered out those with a history of subpar submissions to ensure quality.
Furthermore, workers with <30\% agreement with experts on hidden tests were excluded.
Each batch of work contained annotations for 20 dialogues, taking around 40 minutes to complete, at the cost of £6. Three annotations per annotation task were collected. In case of disagreement, additional annotations were collected until ties were resolved.

\begin{figure}[t]
\centering
\includegraphics[width=1.0\linewidth]{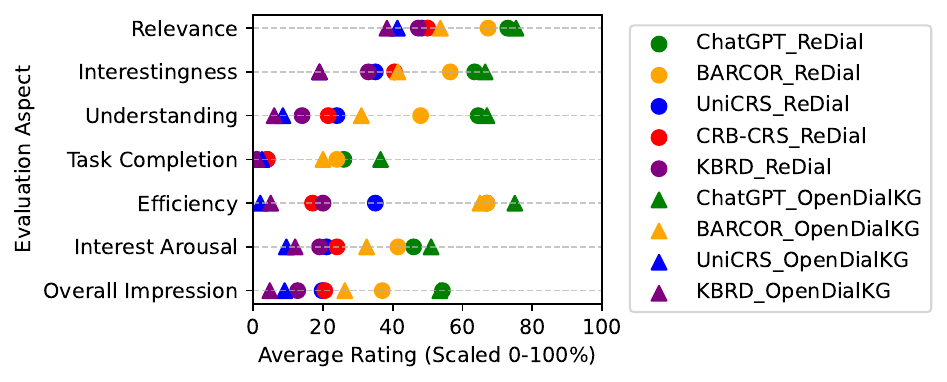}
\shrink
\shrink
\caption{Distribution of human annotation scores for seven aspects across nine systems in CRSArena-Eval.
    The details of seven aspects and nine systems are described in Section~\ref{sec:annotations:process}.
}
\label{fig:crs_distribution}
\end{figure}

\subsection{Analysis}
\label{sec:annotations:analysis}
We now analyze our collected annotations, \emph{CRSArena-Eval}.

\smallskip \noindent \textbf{Statistics.} A total of 20,962 annotations were collected from 109 workers, spanning 467 dialogues and 2,235 system turns.
On average, each dialogue has 14.6 annotation tasks, each annotated by three workers.
Noteworthy, our annotation interface made the process highly efficient, requiring only 8 seconds per annotation.
For 92\% of the tasks, an agreement was achieved by the first three annotators and for the remaining 8\% of tasks additional annotations were collected to resolve ties.
In total, these processes produced 6,805 final labels.

\smallskip \noindent \textbf{Inter-annotator Agreement.}
Given the ordinal nature of judgments, we report inter-annotator agreement using Pearson's $r$ and Spearman's $\rho$, following~\cite{Mehri:2020:USR}, along with Krippendorff's $\alpha$. 
Average scores across all aspects are $r = 0.443$, $\rho = 0.425$, and $\alpha = 0.436$, indicating moderate agreement.
For comparison, the same aspects on the AB-ReDial dataset~\cite{Siro:2022:UUS,Siro:2023:UPU} yield average agreements of $r = 0.328$, $\rho = 0.325$, and $\alpha = 0.327$, showing higher agreement of our CRSArena-Eval annotations.

Table~\ref{tab:agreement} presents the breakdown of inter-annotator agreement for each aspect, showing that CRSArena-Eval consistently achieves higher agreement than AB-ReDial in all aspects except interest arousal.
These results demonstrate that CRSArena-Eval provides higher quality annotations even compared to existing high-quality annotations of the AB-ReDial dataset, showing both the difficulty of the task and high quality annotations of our dataset.

\smallskip \noindent \textbf{System Score Distribution.}
Figure~\ref{fig:crs_distribution} shows the distribution of collected scores for the nine CRSs in CRSArena-Eval.
It shows no system reaches the high end of the scale, indicating that existing CRSs do not fully satisfy users.
This aside, the scores cover a broad range, reflecting differing system quality, which is crucial to assess the ability of automatic evaluators to distinguish system performance~\cite{Mehri:2020:USR}.

\section{Experimental Setup}
\label{sec:setup}

\smallskip \noindent \textbf{Datasets.}
For instruction optimization, we use the  \textit{AB-ReDial}~\cite{Siro:2022:UUS,Siro:2023:UPU} dataset, which contains annotations of human-human conversations from the  ReDial dataset (cf. Sec.~\ref{sec:annotations:process}). 
The annotations are obtained for the seven evaluation aspects and are per turn/dialogue.
This ensures no human-system conversations are involved in the optimization process.
We use 60\% of AB-ReDial for training and the rest for validation.
For evaluation, we use the \textit{CRSArena-Eval} dataset (cf. Sec.~\ref{sec:annotations}).
CRSArena-Eval (RD)/(KG) denote the subset of the dataset for systems developed using ReDial~\citep{Li:2018:TDC} and OpenDialKG~\cite{Moon:2019:OEC}, respectively.

\smallskip \noindent \textbf{Settings.}
Unless indicated otherwise Llama-3.1-8B-Instruct~\cite{Dubey:2024:LHM} is used for FACE.
All experiments were performed using SGLang~\cite{Zheng:2024:SGE} for its inference efficiency. The temperature of 0.6 is set across all experiments unless otherwise stated.
\hi{Following~\cite{Pryzant:2023:APO}}, we run the instruction optimization for $K = 6$ iterations, and stored $b' = 16$ instructions in the instruction pool, resulting in $96$ instructions.
We set parameters $\alpha=2$, $c=1$, $b = 4$,  and use the batch size of $B = |\mathbf{I}'_k|/2$ and $T = 5B$ iterations.
A sampling size of $n = 5$ is used to create a score distribution (Sec.~\ref{sec:method:evaluator}).
All hyperparameters are obtained using the validation set or following~\cite{Pryzant:2023:APO}.
The final selection of the optimal instruction set $\mathbf{I^*}$ (line 18 of Algorithm~\ref{alg:optimization}) is done using the validation set and results in a set of 16 instructions.

\smallskip \noindent \textbf{Prompts.}
Inspired by TREC RAG 2024~\cite{Pradeep:2025:GNR}, the decomposer prompt starts with: \textit{``Your task is to extract conversation nuggets [...]''} followed by CoT prompts and their format.
The textual gradient prompt $\nabla$ begins with \textit{``Examine the original instructions, predicted nugget score, and gold score.''} and then identifies inconsistency between predicted and gold scores, followed by suggestions to the instruction if necessary.
The instruction rewriting prompt $\delta$ starts with \textit{``Propose new instructions of ~50 words based on [...]''} followed by the guidance on how to rewrite based on the textual gradient, inspired by~\cite{Ye:2024:PEP}. 
The seed evaluation instruction $I$ is as follows \textit{``Given the dialogue, evaluate the quality of the target nugget based on the given aspect. Step 1: [...]''}.
Examples of prompts are provided in our repository \href{https://github.com/informagi/face}{\texttt{https://github.com/\allowbreak informagi/\allowbreak face}}.

\smallskip \noindent \textbf{Baselines.}
Multiple automatic evaluators are used as our baselines:
\textit{LLM\textsuperscript{Direct}} directly prompts the LLM to annotate the given turn/dialogue using the same instructions as human annotations for CRSArena-Eval;
\textit{LLM\textsuperscript{CoT+ICL}} adds CoT~\cite{Kojima:2024:LLM,Wei:2022:CTP} and in-context learning (ICL)~\cite{Brown:2020:LMF} with two examples to the prompt of LLM\textsuperscript{Direct}; 
\textit{UniEval}~\cite{Zhong:2022:TUM} and \textit{G-Eval}~\cite{Liu:2023:GNE} are state-of-the-art reference-free conversation evaluation methods.
Since UniEval covers limited aspects, we only report those overlapping with ours.
For fair comparison, we use Llama-3.1-8B-Instruct as the backbone for LLM-based methods LLM\textsuperscript{Direct}, LLM\textsuperscript{CoT+ICL}, and G-Eval in Section~\ref{sec:results:annotation_correlation}.
To show generalizability to different LLMs in Section~\ref{sec:results:generalizability}, other LLMs are used for our experiments.

\smallskip \noindent \textbf{Metrics.}
For correlation metrics, we use Pearson's and Spearman's to appropriately handle the annotation scales. Following~\cite{Mehri:2020:USR}, correlation significance is computed by p-value derived from t-distribution using Python's SciPy library~\cite{Virtanen:2020:FAS}.

\section{Results}
\label{sec:results}
We begin by outlining our key research questions and then present a series of experiments conducted to address them: 
\textbf{RQ1:} How does FACE correlate with human judgments?
\textbf{RQ2:} How generalizable are FACE-optimized instructions across different LLMs and domains?
\textbf{RQ3:} Can fine-grained evaluation scores of FACE provide insights about system's issues?

\begin{table}[t]
\shrink
\centering
\caption{System ranking correlations on CRSArena-Eval, averaged over corresponding aspects.  All FACE correlations are statistically significant with $p < 0.05$.} 
\shrink
\label{tab:ranking_correlation}
\setlength{\tabcolsep}{3.7pt}
\begin{tabular}{l|cc|cc|cc}
\hline
\textbf{Methods} & \multicolumn{2}{c|}{\textbf{Turn-level}} & \multicolumn{2}{c|}{\textbf{Dial-level}} & \multicolumn{2}{c}{\textbf{All}} \\
 & $r$ & $\rho$ & $r$ & $\rho$ & $r$ & $\rho$ \\
\Xhline{2pt}
R@1 & -0.197 & 0.060 & -0.120 & 0.081 & -0.142 & 0.075 \\
R@10 & -0.192 & 0.048 & -0.111 & 0.071 & -0.134 & 0.064 \\
iEvaLM\textsuperscript{free} & 0.191 & 0.191 & 0.232 & 0.232 & 0.257 & 0.184 \\
iEvaLM\textsuperscript{attr} & 0.527 & 0.527 & 0.553 & 0.553 & 0.575 & 0.517 \\
Distinct-3 & 0.716 & 0.841 & 0.665 & 0.780 & 0.680 & 0.798 \\
Distinct-4 & 0.654 & 0.800 & 0.609 & 0.760 & 0.622 & 0.771 \\
LLM\textsuperscript{Direct} & 0.860 & 0.822 & 0.872 & 0.799 & 0.868 & 0.806 \\
G-Eval & 0.740 & 0.840 & 0.893 & 0.830 & 0.850 & 0.833 \\
\hline
FACE & \textbf{0.930} & \textbf{0.842} & \textbf{0.913} & \textbf{0.837} & \textbf{0.918} & \textbf{0.838} \\
\hline
\end{tabular}
\end{table}

\subsection{FACE Annotation Correlation}
\label{sec:results:annotation_correlation}

\smallskip \noindent \textbf{Annotation Correlation.}
Table~\ref{tab:correlation_main} shows the annotation correlation results, demonstrating that FACE, on average, outperforms all baselines by a large margin. 
We note that FACE is not optimized on any human-system conversations (cf. Sec.~\ref{sec:setup}), highlighting its strong generalization to unseen systems.
Although one can argue that FACE can capture some information from human-human ReDial dataset during the instruction optimization process, results on CRSArena-Eval (KG) shows generalization and robustness of FACE to unseen recommendation datasets. 

The results also demonstrate that even without instruction optimization (FACE w/o train), FACE remains competitive with the state-of-the-art method, G-Eval, suggesting the effectiveness of our particle-based approach. 
The benefit of training is more pronounced for challenging aspects such as \textit{Interestingness}, \textit{Efficiency}, and \textit{Interest Arousal}, which indicates that FACE effectively optimizes instructions for aspects that LLMs struggle to capture using a single thought process.

\smallskip \noindent \textbf{Ranking Correlation.}
Table~\ref{tab:ranking_correlation} shows the correlation of system rankings created by different automatic evaluation methods.
System ranking correlations are calculated by averaging each system's score, ranking systems, and measuring correlation with system rankings based on human judgments.
We obtain correlation for reference-based metrics  by computing system rankings from reported scores~\cite{Wang:2023:REC}.
As baselines, we report Recall, and Distinct-n~\cite{Li:2016:DPO} as reference-based metrics, iEvaLM~\cite{Wang:2023:REC} as state-of-the-art CRS specific evaluation method, and LLM\textsuperscript{Direct} and G-Eval as top-performing reference-free metrics from Table~\ref{tab:correlation_main}.
iEvaLM has two variants: iEvaLM\textsuperscript{free} uses LLM-based free-form user simulation, while iEvaLM\textsuperscript{attr} employs attribute-based user simulation~\cite{Wang:2023:REC}.

From the results, we notice that recall metrics are insufficient, which is in line with the literature~\cite{Bernard:2025:CRS}. Surprisingly, iEvaLM baselines struggle to achieve high ranking correlations; this could be due to their reliance on a recall-based approach to evaluate the simulation results.
While the results show that FACE outperforms all baselines, we note that system ranking correlations should be consumed with caution, as they are less informative than annotation correlation, especially for competitive systems~\cite{Faggioli:2023:PLL, dietz:2025:llmtropes}. 

\smallskip \noindent
Overall, we can answer \textbf{(RQ1)}: FACE achieves high annotation and system ranking correlations with human judgments, outperforming state-of-the-art methods by a large margin.

\begin{table}[t]
\shrink
\centering
\caption{Results on generalizability of FACE to other LLMs. 
CRS-RD and -KG represent CRSArena-Eval (RD) and (KG), respectively.
All FACE annotation correlations are statistically significant with $p < 0.01$.}
\shrink
\label{tab:llm_generalizability}
\setlength{\tabcolsep}{1.5pt}
\begin{tabular}{l|l|c||cc|cc|cc}
\hline
\multirow{2}{*}{\textbf{Methods}} & \multirow{2}{*}{\textbf{LLM}} & \multirow{2}{*}{\textbf{Size}} &
\multicolumn{2}{c|}{\textbf{CRS-RD}} &
\multicolumn{2}{c|}{\textbf{CRS-KG}} &
\multicolumn{2}{c}{\textbf{Avg.}} \\
 & & &  $r$ & $\rho$ &  $r$ & $\rho$ & $r$ & $\rho$  \\
\Xhline{2pt}
LLM\textsuperscript{Direct} & Llama & 8B & 0.564 & 0.522 & 0.655 & 0.557 & 0.610 & 0.540 \\
G-Eval & Llama & 8B & 0.577 & 0.577 & 0.656 & 0.642 & 0.617 & 0.610 \\
FACE & Llama & 8B & \textbf{0.712} & 0.668 & \textbf{0.766} & 0.679 & \textbf{0.739} & 0.674 \\
\hline
FACE* & Gemma & 9B & 0.689 & \textbf{0.687} & 0.718 & \textbf{0.703} & 0.704 & \textbf{0.695} \\
FACE* & Gemma & 2B & 0.647 & 0.603 & 0.728 & 0.646 & 0.688 & 0.625 \\
FACE* & Qwen & 7B & 0.698 & 0.664 & 0.764 & 0.693 & 0.731 & 0.679 \\
FACE* & Qwen & 3B & 0.643 & 0.632 & 0.725 & 0.674 & 0.684 & 0.653 \\
FACE* & Qwen & 1.5B & 0.557 & 0.606 & 0.605 & 0.635 & 0.581 & 0.621 \\
\hline
\end{tabular}
\end{table}

\begin{table}[t]
\centering
\caption{Results on generalizability of FACE to chit-chat conversations. All FACE correlations are statistically significant with $p < 0.01$.}
\shrink
\label{tab:domain_generalizability}
\begin{tabular}{l||cc|cc|cc}
\hline
\multirow{2}{*}{\textbf{Methods}} &
\multicolumn{2}{c|}{\small\textbf{USR-Persona}} &
\multicolumn{2}{c|}{\small\textbf{USR-Topical}} &
\multicolumn{2}{c}{\textbf{Avg.}} \\
 &  $r$ & $\rho$ &  $r$ & $\rho$ & $r$ & $\rho$  \\
\Xhline{2pt}
ROUGE-L & 0.114 & 0.091 & 0.193 & 0.203 & 0.154 & 0.147 \\
BLEU-4 & 0.147 & 0.151 & 0.131 & 0.235 & 0.139 & 0.193 \\
METEOR & 0.250 & 0.256 & 0.250 & 0.302 & 0.250 & 0.279 \\
BERTScore & 0.188 & 0.157 & 0.214 & 0.233 & 0.201 & 0.195 \\
Dial-M & 0.400 & 0.390 & 0.370 & 0.400 & 0.385 & 0.395 \\
USR & 0.607 & 0.528 & 0.416 & 0.377 & 0.512 & 0.453 \\
UniEval & 0.616 & 0.580 & \textbf{0.595} & \textbf{0.613} & 0.605 & 0.597 \\
    G-Eval\textsuperscript{GPT-3.5} & 0.441 & 0.458 & 0.519 & 0.544 & 0.480 & 0.501 \\
G-Eval\textsuperscript{GPT-4} & 0.607 & 0.670 & 0.594 & 0.605 & 0.601 & 0.638 \\
\hline
FACE & 0.473 & 0.544 & 0.498 & 0.506 & 0.486 & 0.525 \\
FACE${\text{*}_{\hspace{-0.3em}\text{72B}}}$ & \textbf{0.681} & \textbf{0.697} & 0.570 & 0.582 & \textbf{0.625} & \textbf{0.639} \\
\hline
\end{tabular}
\end{table}

\subsection{Generalizability of FACE}
\label{sec:results:generalizability}
We hypothesize that the pool of optimized instructions by FACE can be reused for different LLMs and domains. To assess this hypothesis, we take the instruction pool and re-select the top instructions (line 18 of Algorithm~\ref{alg:optimization}) for different LLMs and datasets. We denote this adapted FACE as \emph{FACE*}.

\smallskip \noindent \textbf{Generalization to other LLMs.} To assess generalizability of FACE to other LLMs, we use the AB-ReDial validation set and re-select instructions for five LLMs: Gemma 2 (9B and 2B) and Qwen 2.5 (7B, 3B, 1.5B). 
Table~\ref{tab:llm_generalizability} shows annotation correlation results for top-performing baselines of Table~\ref{tab:correlation_main}. The results indicate that adapting to Gemma 9B and Qwen 7B achieves performance comparable to FACE.
Interestingly, our method is highly effective for small models: FACE adapted to Gemma 2B outperforms G-Eval, which uses an LLM with 4x more parameters. A similar observation can be made for Qwen 3B and 1.5B.

\smallskip \noindent \textbf{Generalization to chit-chat conversations.}
To examine generalizability of FACE to another type of conversations, we evaluate FACE on the existing chit-chat datasets: (1) \emph{USR-Persona}~\cite{Mehri:2020:USR} (based on PersonaChat~\cite{Zhang:2018:PDA}), containing personalized chit-chats, and (2) \emph{USR-Topical}~\cite{Mehri:2020:USR} (based on Topical-Chat~\cite{Gopalakrishnan:2019:TTK}), containing knowledge-grounded conversations. These datasets provide annotations for six evaluation aspects, of which ``maintains context'' is the only aspect that is similar to ours. We use the instruction pool for the relevance aspect and re-select optimal instructions using the validation set of USR-Persona for USR-Topical evaluation and vice versa, to ensure that the test set is completely unseen.

Table~\ref{tab:domain_generalizability} presents the results of FACE generalizability to chit-chat conversations, with G-Eval results obtained using GPT-3.5 and 4.
On average, FACE outperforms all baselines except G-Eval with GPT-4, while FACE${\text{*}_{\hspace{-0.3em}\text{72B}}}$ outperforms all baselines.
This is especially striking, considering that FACE is completely blind to category of conversations and uses an LLM with a lower number of parameters than GPT. Additionally, using an open model for evaluation has the added value of reproducibility. 
Based on the results of Tables~\ref{tab:llm_generalizability} and~\ref{tab:domain_generalizability}, we answer our second research question \textbf{(RQ2)}: FACE-optimized instructions are highly generalizable to different LLMs and domains, by performing a simple adaptation of FACE to a new LLM/domain. The adaptation to larger models can even surpass the state-of-the-art method with GPT-4 as a backbone on chit-chat conversations.

\subsection{FACE Interpretability}
\label{sec:results:interpretability}

\begin{figure}[t]
\centering
\shrink
\includegraphics[width=1.0\linewidth]{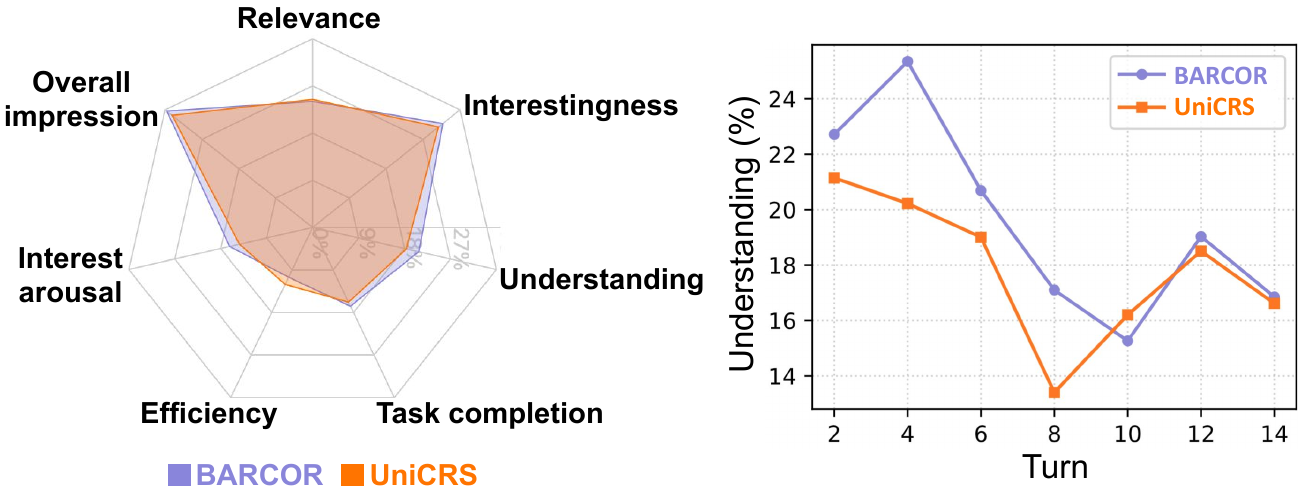}
\shrink
\caption{Breakdown analysis comparing BARCOR and UniCRS.
Left: FACE evaluation results for each aspect. 
Right: Scores of Understanding aspect per system turn.
}
\shrink
\label{fig:interpretability}
\end{figure}

We present a preliminary small-scale case study to demonstrate how FACE fine-grained scores can assist humans in identifying issues in CIA systems. Specifically, we compare FACE scores for two systems that are difficult to diagnose and receive contradictory evaluations from human and existing metrics:
BARCOR~\cite{Wang:2022:BTU} and UniCRS~\cite{Wang:2022:UCR}.
Humans and FACE prefer BARCOR (cf. Fig.~\ref{fig:crs_distribution} and \cite{Bernard:2025:CRS}), while recall-based metrics favor UniCRS~\cite{Wang:2023:REC}.
The left radar chart in Figure~\ref{fig:interpretability} shows FACE analysis for BARCOR and UniCRS for seven evaluation aspects.
Although the overall impression indicates similar performance, FACE scores show that UniCRS excels in relevance and efficiency, while BARCOR is better in user understanding and keeping users interested.
The right graph shows the detailed analysis of the user understanding aspect, where particle scores are aggregated for each turn. It is evident from the plot that BARCOR obtains higher scores in earlier turns. This indicates that BARCOR understands user preferences early on, which may explain its higher human preference.
These insights suggest that, while UniCRS excels in recommendations, overall performance can be improved by focusing on user understanding.

\hi{To further understand each particle's contribution, we analyze particle scores per dialogue act and find that ``preference elicitation'' scored higher for BARCOR than UniCRS (45.5 vs. 39.0; scaled 0-100\%). This demonstrates that BARCOR's superior user understanding stems from more effective preference elicitation, an insight overlooked by item-level recall metrics, which highlights the strength of FACE's interpretability.}

Overall, we answer \textbf{(RQ3)} positively: FACE can provide valuable insights into systems' behavior, which are useful for system improvement.

\begin{figure}[t]
\centering
\shrink
\includegraphics[width=0.93\linewidth]{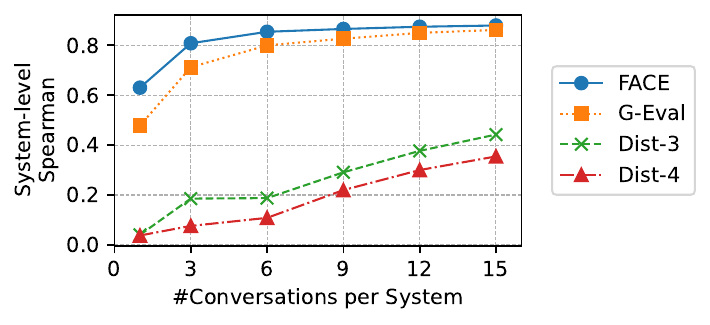}
\shrink
\caption{Sample efficiency of different evaluation methods for the overall aspect on CRSArena-Eval.}
\vspace{-1pt}
\label{fig:sample_efficiency}
\end{figure}

\section{Analysis}
\label{sec:results:analysis}

\textbf{Sample Efficiency.}
To determine system scores and rankings, the evaluation method requires a sample of user-system conversations.
To measure how many samples are needed to find a system ranking with a high correlation with human judgments, we plot system ranking correlations for various conversation counts per system in Figure~\ref{fig:sample_efficiency}.
The results indicate that FACE has strong sample efficiency; it achieves a Spearman correlation of 0.8 with gold rankings using only 3 dialogues per system, making it twice as efficient as the best-performing existing method, G-Eval.
Given that collecting human-system conversations require cost and effort, FACE's sample efficiency significantly enhances actual usability. 

\smallskip \noindent \textbf{Bias Analysis.}
We analyze whether FACE shows known LLM biases: length bias and self-bias (cf. Sec.~\ref{sec:related}).
For \textbf{length bias}~\cite{Dubois:2024:LCA,Wang:2024:HCE}, using CRSArena-Eval, we examine the correlation between a system's average word count in conversations and the overall score.
We find that Pearson's correlations are 0.824 and 0.868 for FACE and humans, respectively. This indicates no sign of length bias compared to humans, which is in line with existing work~\cite{Chiang:2025:CAO,Dubois:2024:LCA} that report humans also favour longer responses, highlighting the nuanced nature of the LLM length bias.

For \textbf{self-bias}, where LLMs prefer system responses over human ones, we use FACE to evaluate pairs of system- and human-generated responses and see if they show any preferences compared to gold human annotators.
We examine two conversation types: USR-Persona for chit-chat, and a combination of CRSArena-Eval and AB-ReDial for a recommendation.
We could not find evidence for self-bias in FACE; e.g., for USR-Persona, FACE aligns with human preferences 77.8\% of the time when humans prefer human-generated responses and 71.4\% of the time when they prefer system-generated responses.

\section{Conclusion}

We present FACE, a fine-grained, aspect-based evaluation method for \hi{conversational information access systems}.
It addresses the shortcomings of existing metrics, such as focusing on fixed dialogue history with reference-based metrics, \hi{limited generalizability of LLM-based metrics}, and relying on non-granular scores with limited insights.
FACE is shown to strongly correlate with human judgments, generalize across LLMs and domains, and provide insights for system improvement.
Future work needs to address current limitations by further examining evaluation biases, assessing effectiveness across broad domains, and exploring how FACE can help expert evaluators.



\balance
\bibliographystyle{ACM-Reference-Format}
\bibliography{references-base}

@article{Trippas2025,
   author = {Johanne R Trippas and J Shane Culpepper and Mohammad Aliannejadi and James Allan and Enrique Amigó and Jaime Arguello and Leif Azzopardi and Peter Bailey and Jamie Callan and Rob Capra and Nick Craswell and Bruce Croft and Jeff Dalton and Gianluca Demartini and Laura Dietz and Zhicheng Dou and Carsten Eickhoff and Michael Ekstrand and Nicola Ferro and Norbert Fuhr and Dorota Glowacka and Faegheh Hasibi and Danula Hettiachchi and Rosie Jones and Jaap Kamps and Noriko Kando and Sarvnaz Karimi and Makoto P Kato and Bevan Koopman and Yiqun Liu and Chenglong Ma and Joel Mackenzie and Maria Maistro and Jiaxin Mao and Dana McKay and Bhaskar Mitra and Stefano Mizzaro and Alistair Moffat and Josiane Mothe and Iadh Ounis and Lida Rashidi and Yongli Ren and Mark Sanderson and Rodrygo Santos and Falk Scholer and Chirag Shah and Laurianne Sitbon and Ian Soboroff and Damiano Spina and Paul Thomas and Julián Urbano and Arjen de Vries and Ryen White and Abby Yuan and Hamed Zamani and Oleg Zendel and Min Zhang and Shengyao Zhuang and Justin Zobel and Guido Zuccon},
   doi = {10.1145/3769733.3769739},
   issn = {0163-5840},
   issue = {1},
   journal = {SIGIR Forum},
   month = {10},
   pages = {1-68},
   title = {Report from the 4th Strategic Workshop on Information Retrieval in Lorne (SWIRL 2025)},
   volume = {59},
   year = {2025}
}

@article{dietz:2025:llmtropes,
  title     = {Principles and Guidelines for the Use of LLM Judges},
  author    = {Dietz, Laura and Zendel, Oleg and Bailey, Peter and Clarke, Charles L. A. and Cotterill, Ellese and Dalton, Jeff and Hasibi, Faegheh and Sanderson, Mark and Craswell, Nick},
  booktitle = {Proceedings of the 2025 International ACM SIGIR Conference on Innovative Concepts and Theories in Information Retrieval},
  year      = {2025}
}

@inproceedings{Yang:2025:LLM,
  author    = {Yang, Chengrun and Wang, Xuezhi and Lu, Yifeng and Liu, Hanxiao and Le, Quoc V and Zhou, Denny and Chen, Xinyun},
  title     = {Large Language Models as Optimizers},
  booktitle = {Proceedings of International Conference on Learning Representations},
  year      = {2025}
}

@inproceedings{Razavi:2025:BPS,
  author    = {Razavi, Amirhossein and Soltangheis, Mina and Arabzadeh, Negar and Salamat, Sara and Zihayat, Morteza and Bagheri, Ebrahim},
  title     = {Benchmarking Prompt Sensitivity in Large Language Models},
  booktitle = {Advances in Information Retrieval: 47th European Conference on Information Retrieval, ECIR 2025, Lucca, Italy, April 6–10, 2025, Proceedings, Part III},
  year      = {2025}
}

@inproceedings{Chen:2024:IEI,
  author    = {Chen, Lichang and Chen, Jiuhai and Goldstein, Tom and Huang, Heng and Zhou, Tianyi},
  title     = {InstructZero: Efficient Instruction Optimization for Black-Box Large Language Models},
  booktitle = {Proceedings of International Conference on Learning Representations},
  year      = {2024}
}

@inproceedings{Yang:2024:LLM,
  author    = {Yang, Chengrun and Wang, Xuezhi and Lu, Yifeng and Liu, Hanxiao and Le, Quoc V. and Zhou, Denny and Chen, Xinyun},
  title     = {Large Language Models as Optimizers},
  booktitle = {Proceedings of International Conference on Learning Representations},
  year      = {2024}
}

@article{Yuksekgonul:2024:TAD,
  title   = {Textgrad: Automatic ``differentiation'' via text},
  author  = {Yuksekgonul, Mert and Bianchi, Federico and Boen, Joseph and Liu, Sheng and Huang, Zhi and Guestrin, Carlos and Zou, James},
  journal = {arXiv preprint arXiv:2406.07496},
  year    = {2024}
}

@inproceedings{Abbasiantaeb:2025:CGE,
  author    = {Abbasiantaeb, Zahra and Lupart, Simon and Azzopardi, Leif and Dalton, Jeffery and Aliannejadi, Mohammad},
  title     = {Conversational Gold: Evaluating Personalized Conversational Search System using Gold Nuggets},
  booktitle = {Proceedings of the 48th International ACM SIGIR Conference on Research and Development in Information Retrieval},
  year      = {2025}
}

@article{Virtanen:2020:FAS,
  title   = {SciPy 1.0: Fundamental Algorithms for Scientific Computing in Python},
  author  = {Virtanen, Pauli and Gommers, Ralf and Oliphant, Travis E. and Haberland, Matt and Reddy, Tyler and Cournapeau, David and Burovski, Evgeni and Peterson, Pearu and Weckesser, Warren and Bright, Jonathan and others},
  journal = {Nature Methods},
  year    = {2020}
}

@inproceedings{Zheng:2024:SGE,
  author    = {Zheng, Lianmin and Yin, Liangsheng and Xie, Zhiqiang and Sun, Chuyue (Livia) and Huang, Jeff and Yu, Cody Hao and Cao, Shiyi and Kozyrakis, Christos and Stoica, Ion and Gonzalez, Joseph and Barrett, Clark and Sheng, Ying},
  title     = {SGLang: Efficient Execution of Structured Language Model Programs},
  booktitle = {The Thirty-Eighth Annual Conference on Neural Information Processing Systems},
  year      = {2024}
}

@article{Manzoor:2022:RBC,
  title   = {Towards retrieval-based conversational recommendation},
  author  = {Manzoor, Ahtsham and Jannach, Dietmar},
  journal = {Information Systems},
  year    = {2022}
}

@inproceedings{Moon:2019:OEC,
  title     = {OpenDialKG: Explainable Conversational Reasoning with Attention-based Walks over Knowledge Graphs},
  author    = {Moon, Seungwhan and Shah, Pararth and Kumar, Anuj and Subba, Rajen},
  booktitle = {Proceedings of the 57th Annual Meeting of the Association for Computational Linguistics},
  year      = {2019}
}

@inproceedings{Li:2018:TDC,
  author    = {Li, Raymond and Kahou, Samira and Schulz, Hannes and Michalski, Vincent and Charlin, Laurent and Pal, Chris},
  title     = {Towards deep conversational recommendations},
  booktitle = {Proceedings of the 32nd International Conference on Neural Information Processing Systems},
  year      = {2018}
}

@inproceedings{Chen:2019:KBR,
  author    = {Chen, Qibin and Lin, Junyang and Zhang, Yichang and Ding, Ming and Cen, Yukuo and Yang, Hongxia and Tang, Jie},
  title     = {Towards Knowledge-Based Recommender Dialog System},
  booktitle = {Proceedings of the 2019 Conference on Empirical Methods in Natural Language Processing and the 9th International Joint Conference on Natural Language Processing},
  year      = {2019}
}

@article{Dubey:2024:LHM,
  title   = {The Llama 3 Herd of Models},
  author  = {Dubey, Abhimanyu and Jauhri, Abhinav and Pandey, Abhinav and Kadian, Abhishek and Al-Dahle, Ahmad and Letman, Aiesha and Mathur, Akhil and Schelten, Alan and Yang, Amy and Fan, Angela and others},
  journal = {arXiv preprint arXiv:2407.21783},
  year    = {2024}
}

@inproceedings{Wang:2022:UCR,
  author    = {Wang, Xiaolei and Zhou, Kun and Wen, Ji-Rong and Zhao, Wayne Xin},
  title     = {Towards Unified Conversational Recommender Systems via Knowledge-Enhanced Prompt Learning},
  booktitle = {Proceedings of the 28th ACM SIGKDD Conference on Knowledge Discovery and Data Mining},
  year      = {2022}
}

@article{Wang:2022:BTU,
  title   = {Barcor: Towards a unified framework for conversational recommendation systems},
  author  = {Wang, Ting-Chun and Su, Shang-Yu and Chen, Yun-Nung},
  journal = {arXiv preprint arXiv:2203.14257},
  year    = {2022}
}

@article{Schulman:2017:PPO,
  title   = {Proximal policy optimization algorithms},
  author  = {Schulman, John and Wolski, Filip and Dhariwal, Prafulla and Radford, Alec and Klimov, Oleg},
  journal = {arXiv preprint arXiv:1707.06347},
  year    = {2017}
}

@inproceedings{Zhou:2023:LLM,
  title     = {Large Language Models are Human-Level Prompt Engineers},
  author    = {Zhou, Yongchao and Muresanu, Andrei Ioan and Han, Ziwen and Paster, Keiran and Pitis, Silviu and Chan, Harris and Ba, Jimmy},
  booktitle = {Proceedings of International Conference on Learning Representations},
  year      = {2023}
}

@inproceedings{Fernando:2025:PSS,
  author    = {Fernando, Chrisantha and Banarse, Dylan and Michalewski, Henryk and Osindero, Simon and Rockt{\"a}schel, Tim},
  title     = {Promptbreeder: self-referential self-improvement via prompt evolution},
  booktitle = {Proceedings of International Conference on Learning Representations},
  year      = {2025}
}

@article{Chen:2024:RPA,
  title   = {RePrompt: Planning by Automatic Prompt Engineering for Large Language Models Agents},
  author  = {Chen, Weizhe and Koenig, Sven and Dilkina, Bistra},
  journal = {arXiv preprint arXiv:2406.11132},
  year    = {2024}
}

@inproceedings{Kong:2024:PPR,
  title     = {PRewrite: Prompt Rewriting with Reinforcement Learning},
  author    = {Kong, Weize and Hombaiah, Spurthi and Zhang, Mingyang and Mei, Qiaozhu and Bendersky, Michael},
  booktitle = {Proceedings of the 62nd Annual Meeting of the Association for Computational Linguistics},
  year      = {2024}
}

@article{Upadhyay:2024:LSR,
  title   = {A Large-Scale Study of Relevance Assessments with Large Language Models: An Initial Look},
  author  = {Upadhyay, Shivani and Pradeep, Ronak and Thakur, Nandan and Campos, Daniel and Craswell, Nick and Soboroff, Ian and Dang, Hoa Trang and Lin, Jimmy},
  journal = {arXiv preprint arXiv:2411.08275},
  year    = {2024}
}

@inproceedings{Xu:2024:PPL,
  author    = {Xu, Wenda and Zhu, Guanglei and Zhao, Xuandong and Pan, Liangming and Li, Lei and Wang, William},
  title     = {Pride and Prejudice: {LLM} Amplifies Self-Bias in Self-Refinement},
  booktitle = {Proceedings of the 62nd Annual Meeting of the Association for Computational Linguistics (Volume 1: Long Papers)},
  year      = {2024}
}

@inproceedings{Dietz:2024:WAR,
  author    = {Dietz, Laura},
  title     = {A Workbench for Autograding Retrieve/Generate Systems},
  booktitle = {Proceedings of the 47th International ACM SIGIR Conference on Research and Development in Information Retrieval},
  year      = {2024}
}

@inproceedings{Rajput:2011:NTC,
  author    = {Rajput, Shahzad and Pavlu, Virgil and Golbus, Peter B. and Aslam, Javed A.},
  title     = {A nugget-based test collection construction paradigm},
  booktitle = {Proceedings of the 20th ACM International Conference on Information and Knowledge Management},
  year      = {2011}
}

@inproceedings{Takehi:2023:ODQ,
  author    = {Takehi, Rikiya and Watanabe, Akihisa and Sakai, Tetsuya},
  title     = {Open-Domain Dialogue Quality Evaluation: Deriving Nugget-level Scores from Turn-level Scores},
  booktitle = {Proceedings of the Annual International ACM SIGIR Conference on Research and Development in Information Retrieval in the Asia Pacific Region},
  year      = {2023}
}

@inproceedings{Faggioli:2023:PLL,
  author    = {Faggioli, Guglielmo and Dietz, Laura and Clarke, Charles L. A. and Demartini, Gianluca and Hagen, Matthias and Hauff, Claudia and Kando, Noriko and Kanoulas, Evangelos and Potthast, Martin and Stein, Benno and Wachsmuth, Henning},
  title     = {Perspectives on large language models for relevance judgment},
  booktitle = {Proceedings of the 2023 ACM SIGIR International Conference on Theory of Information Retrieval},
  year      = {2023}
}

@inproceedings{Wang:2024:HCE,
  author    = {Wang, Yizhong and Ivison, Hamish and Dasigi, Pradeep and Hessel, Jack and Khot, Tushar and Chandu, Khyathi Raghavi and Wadden, David and MacMillan, Kelsey and Smith, Noah A. and Beltagy, Iz and Hajishirzi, Hannaneh},
  title     = {How far can camels go? exploring the state of instruction tuning on open resources},
  booktitle = {Proceedings of the 37th International Conference on Neural Information Processing Systems},
  year      = {2024}
}

@inproceedings{Ekstrand:2013:ESN,
  author    = {Ekstrand-Abueg, Matthew and Pavlu, Virgil and Kato, Makoto and Sakai, Tetsuya and Yamamoto, Takehiro and Iwata, Mayu},
  title     = {Exploring Semi-Automatic Nugget Extraction for Japanese One Click Access Evaluation},
  booktitle = {Proceedings of the 36th International ACM SIGIR Conference on Research and Development in Information Retrieval},
  year      = {2013}
}

@inproceedings{Lin:2005:AEA,
  author    = {Lin, Jimmy and Demner-Fushman, Dina},
  title     = {Automatically Evaluating Answers to Definition Questions},
  booktitle = {Proceedings of Human Language Technology Conference and Conference on Empirical Methods in Natural Language Processing},
  year      = {2005}
}

@inproceedings{Chiang:2025:CAO,
  author    = {Chiang, Wei-Lin and Zheng, Lianmin and Sheng, Ying and Angelopoulos, Anastasios N. and Li, Tianle and Li, Dacheng and Zhu, Banghua and Zhang, Hao and Jordan, Michael I. and Gonzalez, Joseph E. and Stoica, Ion},
  title     = {Chatbot Arena: An Open Platform for Evaluating {LLMs} by Human Preference},
  booktitle = {Proceedings of International Conference on Learning Representations},
  year      = {2025}
}

@inproceedings{Zhang:2020:BET,
  author    = {Tianyi Zhang and Varsha Kishore and Felix Wu and Kilian Q. Weinberger and Yoav Artzi},
  title     = {BERTScore: Evaluating Text Generation with BERT},
  booktitle = {Proceedings of International Conference on Learning Representations},
  year      = {2020}
}

@inproceedings{Lin:2004:RPA,
  author    = {Lin, Chin-Yew},
  title     = {ROUGE: A Package for Automatic Evaluation of Summaries},
  booktitle = {Text Summarization Branches Out},
  year      = {2004}
}

@inproceedings{Papineni:2002:BMA,
  author    = {Papineni, Kishore and Roukos, Salim and Ward, Todd and Zhu, Wei-Jing},
  title     = {Bleu: a Method for Automatic Evaluation of Machine Translation},
  booktitle = {Proceedings of the 40th Annual Meeting of the Association for Computational Linguistics},
  year      = {2002}
}

@inproceedings{Wang:2023:SIC,
  author    = {Wang, Xuezhi and Wei, Jason and Schuurmans, Dale and Le, Quoc V and Chi, Ed H. and Narang, Sharan and Chowdhery, Aakanksha and Zhou, Denny},
  title     = {Self-Consistency Improves Chain of Thought Reasoning in Language Models},
  booktitle = {Proceedings of International Conference on Learning Representations},
  year      = {2023}
}

@inproceedings{Gopalakrishnan:2019:TTK,
  author    = {Gopalakrishnan, Karthik and Hedayatnia, Behnam and Chen, Qinlang and Gottardi, Anna and Kwatra, Sanjeev and Venkatesh, Anu and Gabriel, Raefer and Hakkani-Tur, Dilek and Amazon Alexa AI},
  title     = {Topical-Chat: Towards Knowledge-Grounded Open-Domain Conversations},
  booktitle = {Proceedings of Interspeech 2019},
  year      = {2019}
}

@inproceedings{Sclar:2024:QLM,
  title     = {Quantifying Language Models' Sensitivity to Spurious Features in Prompt Design or: How I Learned to Start Worrying About Prompt Formatting},
  author    = {Sclar, Melanie and Choi, Yejin and Tsvetkov, Yulia and Suhr, Alane},
  booktitle = {Proceedings of International Conference on Learning Representations},
  year      = {2024}
}

@inproceedings{Voorhees:2003:OTQ,
  author    = {Voorhees, Ellen M.},
  title     = {Overview of the TREC 2003 Question Answering Track},
  booktitle = {Proceedings of the Twelfth Text REtrieval Conference},
  year      = {2003}
}

@inproceedings{Dey:2023:DMF,
  author    = {Dey, Suvodip and Desarkar, Maunendra Sankar},
  title     = {Dial-M: A Masking-based Framework for Dialogue Evaluation},
  booktitle = {Proceedings of the 24th Annual Meeting of the Special Interest Group on Discourse and Dialogue},
  year      = {2023}
}

@inproceedings{Zheng:2024:JLM,
  author    = {Zheng, Lianmin and Chiang, Wei-Lin and Sheng, Ying and Zhuang, Siyuan and Wu, Zhanghao and Zhuang, Yonghao and Lin, Zi and Li, Zhuohan and Li, Dacheng and Xing, Eric P. and Zhang, Hao and Gonzalez, Joseph E. and Stoica, Ion},
  title     = {Judging {LLM-as-a-judge with MT-bench} and Chatbot Arena},
  booktitle = {Proceedings of the 37th International Conference on Neural Information Processing Systems},
  year      = {2024}
}

@inproceedings{Dubois:2024:LCA,
  author    = {Dubois, Yann and Liang, Percy and Hashimoto, Tatsunori B.},
  title     = {Length-Controlled AlpacaEval: A Simple Way to Debias Automatic Evaluators},
  booktitle = {Conference on Language Modeling (COLM)},
  year      = {2024}
}

@inproceedings{Wang:2023:REC,
  title     = {Rethinking the Evaluation for Conversational Recommendation in the Era of Large Language Models},
  author    = {Wang, Xiaolei  and Tang, Xinyu  and Zhao, Xin  and Wang, Jingyuan  and Wen, Ji-Rong},
  year      = 2023,
  booktitle = {Proceedings of the 2023 Conference on Empirical Methods in Natural Language Processing}
}

@inproceedings{Liu:2016:HNE,
  author    = {Liu, Chia-Wei and Lowe, Ryan and Serban, Iulian and Noseworthy, Mike and Charlin, Laurent and Pineau, Joelle},
  title     = {How {NOT} To Evaluate Your Dialogue System: An Empirical Study of Unsupervised Evaluation Metrics for Dialogue Response Generation},
  booktitle = {Proceedings of the 2016 Conference on Empirical Methods in Natural Language Processing},
  year      = {2016}
}

@inproceedings{Bernard:2025:CRS,
  author    = {Bernard, Nolwenn and Joko, Hideaki and Hasibi, Faegheh and Balog, Krisztian},
  title     = {CRS Arena: Crowdsourced Benchmarking of Conversational Recommender Systems},
  booktitle = {Proceedings of the 18th ACM International Conference on Web Search and Data Mining},
  year      = {2025}
}

@inproceedings{Pradeep:2025:GNR,
  title     = {The Great Nugget Recall: Automating Fact Extraction and RAG Evaluation with Large Language Models},
  author    = {Pradeep, Ronak and Thakur, Nandan and Upadhyay, Shivani and Campos, Daniel and Craswell, Nick and Soboroff, Ian and Dang, Hoa Trang and Lin, Jimmy},
  booktitle = {Proceedings of the 48th International ACM SIGIR Conference on Research and Development in Information Retrieval (SIGIR '25)},
  year      = {2025}
}

@inproceedings{Mayfield:2024:EMR,
  author    = {Mayfield, James and Yang, Eugene and Lawrie, Dawn and MacAvaney, Sean and McNamee, Paul and Oard, Douglas W. and Soldaini, Luca and Soboroff, Ian and Weller, Orion and Kayi, Efsun and Sanders, Kate and Mason, Marc and Hibbler, Noah},
  title     = {On the Evaluation of Machine-Generated Reports},
  booktitle = {Proceedings of the 47th International ACM SIGIR Conference on Research and Development in Information Retrieval},
  year      = {2024}
}

@article{Siro:2023:UPU,
  title   = {Understanding and Predicting User Satisfaction with Conversational Recommender Systems},
  author  = {Siro, Clemencia and Aliannejadi, Mohammad and De Rijke, Maarten},
  journal = {ACM Trans. Inf. Syst.},
  year    = {2023}
}

@inproceedings{Eickhoff:2011:HCY,
  title     = {How Crowdsourcable is Your Task?},
  author    = {Eickhoff, Carsten and de Vries, Arjen P.},
  booktitle = {Proc. of CSDM '11},
  year      = 2011,
  pages     = {11--14}
}

@article{Sakai:2023:SGF,
  title   = {{SWAN}: A Generic Framework for Auditing Textual Conversational Systems},
  author  = {Sakai, Tetsuya},
  journal = {arXiv preprint arXiv:2305.08290},
  year    = {2023}
}

@inproceedings{Siro:2022:UUS,
  author    = {Siro, Clemencia and Aliannejadi, Mohammad and de Rijke, Maarten},
  title     = {Understanding User Satisfaction with Task-oriented Dialogue Systems},
  booktitle = {Proceedings of the 45th International ACM SIGIR Conference on Research and Development in Information Retrieval},
  year      = {2022}
}

@inproceedings{Ye:2024:PEP,
  title     = {Prompt Engineering a Prompt Engineer},
  author    = {Ye, Qinyuan and Ahmed, Mohamed and Pryzant, Reid and Khani, Fereshte},
  booktitle = {Findings of the Association for Computational Linguistics ACL 2024},
  year      = {2024}
}

@inproceedings{Pryzant:2023:APO,
  author    = {Pryzant, Reid and Iter, Dan and Li, Jerry and Lee, Yin and Zhu, Chenguang and Zeng, Michael},
  title     = {Automatic Prompt Optimization with ``Gradient Descent'' and Beam Search},
  booktitle = {Proceedings of the 2023 Conference on Empirical Methods in Natural Language Processing},
  year      = {2023}
}

@inproceedings{Kojima:2024:LLM,
  author    = {Kojima, Takeshi and Gu, Shixiang Shane and Reid, Machel and Matsuo, Yutaka and Iwasawa, Yusuke},
  title     = {Large language models are zero-shot reasoners},
  booktitle = {Proceedings of the 36th International Conference on Neural Information Processing Systems},
  year      = {2024}
}

@article{Alaofi:2024:GIR,
  title   = {Generative Information Retrieval Evaluation},
  author  = {Alaofi, Marwah and Arabzadeh, Negar and Clarke, Charles LA and Sanderson, Mark},
  journal = {arXiv preprint arXiv:2404.08137},
  year    = {2024}
}

@article{Upadhyay:2024:LCP,
  title   = {{LLMs} Can Patch Up Missing Relevance judgements in Evaluation},
  author  = {Upadhyay, Shivani and Kamalloo, Ehsan and Lin, Jimmy},
  journal = {arXiv preprint arXiv:2405.04727},
  year    = {2024}
}

@inproceedings{Joko:2024:DPL,
  title     = {Doing Personal LAPS: {LLM}-Augmented Dialogue Construction for Personalized Multi-Session Conversational Search},
  author    = {Joko, Hideaki and Chatterjee, Shubham and Ramsay, Andrew and de Vries, Arjen P. and Dalton, Jeff and Hasibi, Faegheh},
  booktitle = {Proceedings of the 47th International ACM SIGIR Conference on Research and Development in Information Retrieval},
  year      = {2024}
}

@article{soudani2024survey,
  title   = {A survey on recent advances in conversational data generation},
  author  = {Soudani, Heydar and Petcu, Roxana and Kanoulas, Evangelos and Hasibi, Faegheh},
  year = {2026},
  journal = {ACM Comput. Surv.}
}

@article{Aliannejadi:2024:IKAT,
  title   = {TREC iKAT 2023: The Interactive Knowledge Assistance Track Overview},
  author  = {Aliannejadi, Mohammad and Abbasiantaeb, Zahra and Chatterjee, Shubham and Dalton, Jeffery and Azzopardi, Leif},
  journal = {arXiv preprint arXiv:2401.01330},
  year    = {2024}
}

@inproceedings{Wei:2022:CTP,
  author    = {Wei, Jason and Wang, Xuezhi and Schuurmans, Dale and Bosma, Maarten and ichter, brian and Xia, Fei and Chi, Ed and Le, Quoc V and Zhou, Denny},
  title     = {Chain-of-Thought Prompting Elicits Reasoning in Large Language Models},
  booktitle = {Advances in Neural Information Processing Systems},
  year      = {2022}
}

@inproceedings{Zhong:2022:TUM,
  title     = {Towards a Unified Multi-Dimensional Evaluator for Text Generation},
  author    = {Zhong, Ming and Liu, Yang and Yin, Da and Mao, Yuning and Jiao, Yizhu and Liu, Pengfei and Zhu, Chenguang and Ji, Heng and Han, Jiawei},
  booktitle = {Proceedings of the 2022 Conference on Empirical Methods in Natural Language Processing},
  year      = {2022}
}

@inproceedings{Mehri:2020:USR,
  author    = {Mehri, Shikib and Eskenazi, Maxine},
  title     = {{USR}: An Unsupervised and Reference Free Evaluation Metric for Dialog Generation},
  booktitle = {Proceedings of the 58th Annual Meeting of the Association for Computational Linguistics},
  year      = {2020}
}

@inproceedings{Liu:2023:GNE,
  author    = {Liu, Yang and Iter, Dan and Xu, Yichong and Wang, Shuohang and Xu, Ruochen and Zhu, Chenguang},
  title     = {G-Eval: NLG Evaluation using Gpt-4 with Better Human Alignment},
  booktitle = {Proceedings of the 2023 Conference on Empirical Methods in Natural Language Processing},
  year      = {2023}
}

@inproceedings{Brown:2020:LMF,
  title     = {Language Models are Few-Shot Learners},
  author    = {Brown, Tom and Mann, Benjamin and Ryder, Nick and Subbiah, Melanie and Kaplan, Jared D and Dhariwal, Prafulla and Neelakantan, Arvind and Shyam, Pranav and Sastry, Girish and Askell, Amanda and Agarwal, Sandhini and Herbert-Voss, Ariel and Krueger, Gretchen and Henighan, Tom and Child, Rewon and Ramesh, Aditya and Ziegler, Daniel and Wu, Jeffrey and Winter, Clemens and Hesse, Chris and Chen, Mark and Sigler, Eric and Litwin, Mateusz and Gray, Scott and Chess, Benjamin and Clark, Jack and Berner, Christopher and McCandlish, Sam and Radford, Alec and Sutskever, Ilya and Amodei, Dario},
  booktitle = {Advances in Neural Information Processing Systems},
  year      = {2020}
}

@inproceedings{Bernard:2023:MGS,
  title     = {MG-ShopDial: A Multi-Goal Conversational Dataset for e-Commerce},
  author    = {Bernard, Nolwenn and Balog, Krisztian},
  booktitle = {Proceedings of the 46th International ACM SIGIR Conference on Research and Development in Information Retrieval, July 23--27, 2023, Taipei, Taiwan},
  year      = {2023}
}

@inproceedings{Li:2016:DPO,
  title     = {A Diversity-Promoting Objective Function for Neural Conversation Models},
  author    = {Li, Jiwei and Galley, Michel and Brockett, Chris and Gao, Jianfeng and Dolan, William B},
  booktitle = {Proceedings of the 2016 Conference of the North American Chapter of the Association for Computational Linguistics: Human Language Technologies},
  year      = {2016}
}

@inproceedings{Lin:2023:UMA,
  title     = {{LLM-Eval}: Unified Multi-Dimensional Automatic Evaluation for Open-Domain Conversations with Large Language Models},
  author    = {Lin, Yen-Ting and Chen, Yun-Nung},
  booktitle = {Proceedings of the 5th Workshop on NLP for Conversational AI},
  year      = {2023}
}

@inproceedings{Joko:2021:CEL,
  author    = {Joko, Hideaki and Hasibi, Faegheh and Balog, Krisztian and de Vries, Arjen P.},
  title     = {Conversational Entity Linking: Problem Definition and Datasets},
  booktitle = {Proceedings of the 44rd International ACM SIGIR Conference on Research and Development in Information Retrieval},
  year      = {2021}
}

@inproceedings{Zhang:2018:PDA,
  title     = {Personalizing Dialogue Agents: {I} have a dog, do you have pets too?},
  author    = {Zhang, Saizheng  and
               Dinan, Emily  and
               Urbanek, Jack  and
               Szlam, Arthur  and
               Kiela, Douwe  and
               Weston, Jason},
  booktitle = {Proceedings of the 56th Annual Meeting of the Association for Computational Linguistics},
  year      = {2018}
}

@inproceedings{Balog:2025:RJA,
  title     = {Rankers, Judges, and Assistants: Towards Understanding the Interplay of LLMs in Information Retrieval Evaluation},
  author    = {Balog, Krisztian and Metzler, Donald and Qin, Zhen},
  booktitle = {Proceedings of the 48th International ACM SIGIR Conference on Research and Development in Information Retrieval},
  year      = {2025}
}

@inproceedings{Bernard:2025:LCE,
  title     = {Limitations of Current Evaluation Practices for Conversational Recommender Systems and the Potential of User Simulation},
  author    = {Bernard, Nolwenn and Balog, Krisztian},
  booktitle = {Proceedings of the 2025 International ACM SIGIR Conference on Research and Development in Information Retrieval in the Asia Pacific Region},
  year      = {2025}
}

@inproceedings{Farzi:2024:ELA,
  title     = {Exam++: LLM-based Answerability Metrics for IR Evaluation},
  author    = {Farzi, Naghmeh and Dietz, Laura},
  booktitle = {Proceedings of LLM4Eval: The First Workshop on Large Language Models for Evaluation in Information Retrieval},
  year      = {2024}
}

@inproceedings{Aliannejadi:2025:iKAT,
  title     = {TREC iKAT 2025: The Interactive Knowledge Assistance Track Overview},
  author    = {Aliannejadi, Mohammad and Lupart, Simon and Gohsen, Marcel and Abbasiantaeb, Zahra and Mirzakhmedova, Nailia and Kiesel, Johannes and Dalton, Jeffrey},
  booktitle = {Proceedings of the 34th Text REtrieval Conference (TREC 2025)},
  year      = {2025}
}

@inproceedings{Thakur:2025:AST,
  title     = {Assessing Support for the TREC 2024 RAG Track: A Large-Scale Comparative Study of LLM and Human Evaluations},
  author    = {Thakur, Nandan and Pradeep, Ronak and Upadhyay, Shivani and Campos, Daniel and Craswell, Nick and Soboroff, Ian and Dang, Hoa Trang and Lin, Jimmy},
  booktitle = {Proceedings of the 48th International ACM SIGIR Conference on Research and Development in Information Retrieval (SIGIR '25)},
  year      = {2025}
}

@inproceedings{Yu:2025:BPS,
  title     = {Beyond Pointwise Scores: Decomposed Criteria-Based Evaluation of {LLM} Responses},
  author    = {Yu, Fangyi and Seedat, Nabeel and Herrmannova, Drahomira and Schilder, Frank and Schwarz, Jonathan Richard},
  booktitle = {Proceedings of the 2025 Conference on Empirical Methods in Natural Language Processing: Industry Track},
  year      = {2025}
}

@inproceedings{Joko:2026:WIA,
  title     = {WildClaims: Information Access Conversations in the Wild (Chat)},
  author    = {Joko, Hideaki and Amirshahi, Shakiba and Clarke, Charles L. A. and Hasibi, Faegheh},
  booktitle = {Proceedings of the 48th European Conference on Information Retrieval (ECIR 2026)},
  year      = {2026}
}

\end{document}